%% file: COCrates.tex
\documentclass[smallextended]{svjour3}  
\usepackage{booktabs}  
\usepackage{graphicx}  
\usepackage{natbib}
\usepackage{etex}	
\usepackage{acronym}  
\usepackage{amsmath}	
\usepackage{amssymb}	
\usepackage{ulem}
\usepackage[colorlinks=true,citecolor=blue,urlcolor=blue,breaklinks]{hyperref}
\usepackage{caption}

\usepackage{xspace}
\newcommand{\Gpcyr}{\ensuremath{\,\rm{Gpc}^{-3}\,\rm{yr}^{-1}}\xspace}

\journalname{Living Reviews in Relativity}

\begin{document}

\title{Rates of compact object coalescences}

\author{Ilya Mandel \and Floor S.~Broekgaarden}

\institute{I. Mandel \at
Monash Centre for Astrophysics, School of Physics and Astronomy,\\
Monash University, Clayton, Victoria 3800, Australia\\
and\\
OzGrav, ARC Centre of Excellence for Gravitational Wave Discovery, Australia\\
\email{ilya.mandel@monash.edu}\\
\\
F. S. Broekgaarden \at
Center for Astrophysics / Harvard \& Smithsonian,
60 Garden St., Cambridge, MA 02138, USA\\
\email{floor.broekgaarden@cfa.harvard.edu}}

\date{}
\maketitle

\begin{abstract}
Gravitational-wave detections are enabling measurements of the rate of coalescences of binaries composed of two compact objects -- neutron stars and/or black holes.  The coalescence rate of binaries containing neutron stars is further constrained by electromagnetic observations, including Galactic radio binary pulsars and short gamma-ray bursts.  Meanwhile, increasingly sophisticated models of compact objects merging through a variety of evolutionary channels produce a range of theoretically predicted rates.  Rapid improvements in instrument sensitivity, along with plans for new and improved surveys, make this an opportune time to summarise the existing observational and theoretical knowledge of compact-binary coalescence rates. \\

\textbf{Note that this ArXiv version uses v7 of the data, whereas the published article uses v5. For details, see supplementary material on Zenodo, \citet{ZenodoReview:2021}.}
\end{abstract}

\setcounter{tocdepth}{3}
\tableofcontents



\section{Introduction}
\label{sec:intro}

It has been more than a decade since  \citet{ratesdoc} reviewed the literature on compact-object merger rate predictions.  At that time, a handful of ultimately merging double neutron star (NS) systems in the Milky Way Galaxy were known from radio pulsar observations, starting with the discovery of \citet{HulseTaylor:1975}.  It was believed that short gamma-ray bursts (SGRBs) were associated with mergers of two NSs.  But no actual compact object mergers were definitively observed.  
Therefore, most of the rate predictions, particularly for mergers involving black holes (BHs), were based purely on theoretical models (see, e.g., \citealt{MandelOShaughnessy:2010} for a review of the status of models at that time).

The intervening decade has completely changed this landscape.  On 14 September, 2015, the advanced Laser Interferometer Gravitational-wave Observatory (LIGO, \citealt{AdvLIGO}) detected the first chirp of gravitational waves from a binary BH merger, GW150914 \citep{GW150914}.  Since then, nine more binary BH coalescences were observed during the first and second observing runs of advanced LIGO and Virgo detectors \citep{BBH:O2}.  The data release from the third observing run brought the total number of confidently detected binary BH coalescences to approximately 70 \citep{GWTC3}.  The first detection of a binary neutron star merger, GW170817 \citep{GW170817}, was accompanied by a short GRB \citep{GW170817:GRB} and a kilonova \citep{GW170817:MMA}, firmly establishing the connection between these phenomena.  A second confident binary neutron star detection, GW190425 \citep{GW190425}, followed, along with detections of two NS--BH mergers,  GW200105 and GW200115 \citep{GW200105}.

At the same time, significant progress has been made in the theoretical modelling of the sources of compact binary coalescences.  This has included more detailed exploration of the classical channel of isolated binary evolution, typically through the common-envelope phase \citep{vdH:1976,SmarrBlandford:1976,TutukovYungelson:1993} (or possibly through stable mass transfer alone, \citealt{vdH:2017,Inayoshi:2017}), along with significant new developments in the investigation of alternative channels, including dynamical interaction in dense stellar environments \citep{Sigurdsson:1993,Kulkarni:1993,PortegiesZwart:2000}; mergers in binaries of rapidly rotating stars undergoing chemically homogeneous evolution \citep{MandelDeMink:2016,Marchant:2016}; mergers facilitated by the Lidov-Kozai resonance \citep{Lidov:1962,Kozai:1962} in hierarchical triple systems; and even mergers of primordial black holes of cosmological rather than astrophysical origin \citep{Bird:2016,AliHaimoud:2017}. 

The coupled advances in observations and theory are making it possible to quantitatively compare rate predictions against observations for the first time \citep[e.g.,][]{BBH:O1O2,GWTC3:pop,Belczynski:2017,Wysocki:2017,Neijssel:2019,Belczynski:2020,Farmer:2020,Bavera:2020}.  Beyond this, they provide critical input into understanding the chemical enrichment of the Universe, particularly with elements created through r-process nucleosynthesis likely associated with mergers involving NSs  with their neutron-rich material \citep{Kasen:2017, Cote:2018, Metzger:2019kilonova, Kobayashi:2020}.  The merger rates determine prospects for future Earth-based and space-borne gravitational-wave detectors, and will impact detector design \citep{,LISA:2017,Adhikari:2019,CE:2019}.   Merger rates feed into constraints or predictions on other observables, ranging from kilonovae associated with binary NS mergers \citep[e.g.,][]{Kasliwal:2020,Mochkovitch:2021} to gravitational-wave stochastic backgrounds \citep{GW150914:stoch} to, possibly, fast radio bursts \citep[e.g.,][]{Zhang:2020}.  This, then, is an opportune time for a review of the current state of the observations and predictions of compact object merger rates -- and, given the expectation of continuing rapid development in this field, it particularly calls for a Living Review. 

This review is organised as follows.  We provide an executive summary of the observed and theoretically predicted compact-binary merger rates  in Sect.~\ref{sec:summary}.  We then provide more detailed information and discussion of the observations in Sect.~\ref{sec:observation} and of the models in  Sect.~\ref{sec:theory} before concluding in Sect.~\ref{sec:outlook}. The collated data and code to reproduce all tables and figures in this review are publicly available through \citet{ZenodoReview:2021}\footnote{The current paper is based on version 6 of the data.}, where we also provide supplementary information on how we obtained the data in the tables and figures.


\section{Executive summary}
\label{sec:summary}

We summarise the coalescence rates in three tables below: NS-NS binaries in Table~\ref{table:BNS1}, NS-BH binaries (where we do not distinguish whether the BH or NS formed first) in Table~\ref{table:NSBH}, and BH-BH binaries in table \ref{table:BBH1}.

Coalescence rates are, in general, functions of redshift; we quote current local rates at redshift $z=0$ per unit source time per unit comoving volume in units of Gpc$^{-3}$ yr$^{-1}$, but caution that these could be up to an order of magnitude larger at higher redshifts.  Where initially stated in different units, we convert these, using, as appropriate, factors of $1.7\times 10^{10}$ solar blue-light luminosities per Milky Way equivalent galaxy (MWEG), $1.17 \times 10^{-2}$ MWEG per Mpc$^3$ \citep{LIGOS3S4Galaxies}, a globular cluster space density of 2.9 per Mpc$^3$ at $z=0$ \citep{PortegiesZwart:2000} and a local supernova rate of  $1.06 \times 10^5\,\rm{Gpc}^{-3}\, \rm{yr}^{-1}$  \citep{Taylor:2014}.  Of course, such simple re-scalings do not account for the dependence of merger rates on the star formation history and metallicity \citep{pacheco:2005,Belczynski:2010,Dvorkin:2016,Neijssel:2019,ChruslinskaNelemans:2019,Mapelli:2021}.

In general, we follow original papers in stating the rates; a discussion follows in the next two sections.  Where uncertainties are stated, we try to indicate them with $\pm$ subscripts and superscripts; however, these are not always available, and may not be attributable to a specific confidence interval when available.  Sometimes, we state a range in square brackets without a central value.

The vast literature on compact-object merger rate predictions dating back to at least 1979 \citep{Clark:1979} makes a complete historical review implausible.  Therefore, we focus on the latest contributions from each group, except where contributions use significantly different methodology or examine the impact of different assumptions.  We generally eschew papers published before 2010 as these have typically been superseded by observational and theoretical advances. 

We combine rates into several broad categories for ease of viewing, but multiple different channels with a broad range of physics may be at play within any one category.  For example,  the category of ``nuclear star clusters'' may include dynamical captures in nuclear clusters without a massive black hole, mergers in hierarchical triples involving a massive black hole, or mergers aided by an accretion disk in an active galactic nucleus.  We do not include some very rare channels with predicted rates below the range shown in the figures.

We do not quote predictions for the number of detections per year for a given instrument because such predictions must take into account the variation in both the merger rate and the mass distribution of merging binaries with redshift for detectors sensitive to cosmological distances, and this information is frequently not readily available in the literature.  Derivations of how to do such calculations are provided by, e.g., \citet{Belczynski:2014VMS} and \citet{deMinkMandel:2016}; both analytical fits (e.g., \citealt{Fishbach:2017mass}) and numerical codes (e.g., \citealt{Gerosa:2017,COMPAS:2021}) are available.


\input{table-BNS.tex}

\begin{figure*}
    \centering
\includegraphics[width=1\textwidth]{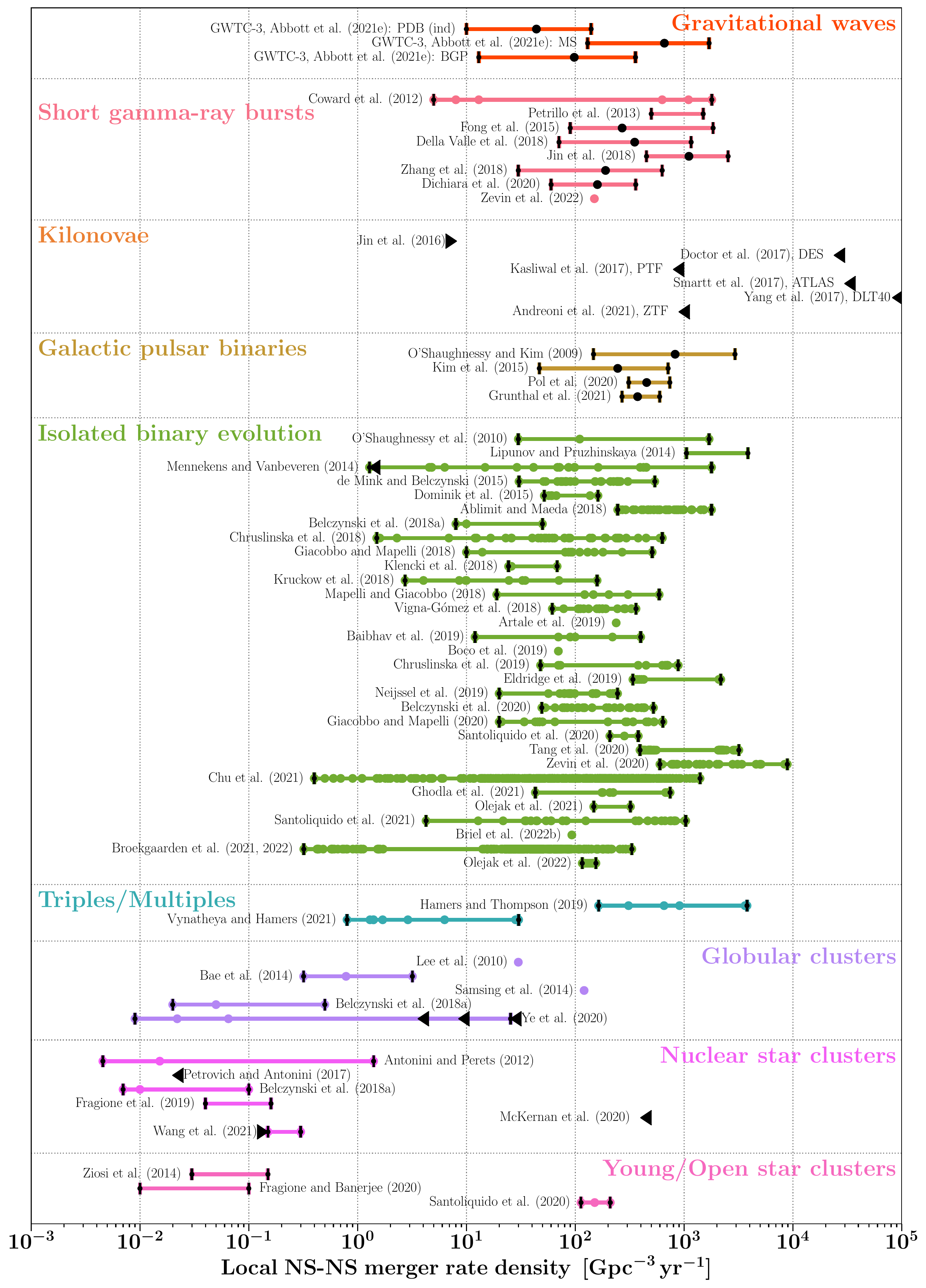} 
    \caption{NS-NS merger rates from Table~\ref{table:BNS1}. Black lines denote the interval boundaries. Triangles mark upper and lower limits. Colored circles mark inferred or simulated values, where multiple values are given in a study.  Black circles mark the median or mean of a confidence interval, where one is provided.
(\href{https://github.com/FloorBroekgaarden/Rates_of_Compact_Object_Coalescence/blob/main/plottingCode/Make_figures_Mandel_and_Broekgaarden_2021_COC_rates_review.ipynb}{GitHub})
    }
    \label{fig:NSNS}
\end{figure*}
%


\section{Observed rates}
\label{sec:observation}

In this section, we discuss the latest observational constraints on compact-object coalescence rates.  There are direct observations of compact binary mergers, either through gravitational waves or through  electromagnetic transients accompanying such mergers, particularly short gamma-ray bursts. Alternatively, there are observations of binaries that are not yet merging, but are expected to do so as compact objects in the future, such as Galactic double neutron stars and perhaps some high-mass X-ray binaries.  

In all cases the key challenge in accurately inferring rates from limited observations comes from significant selection effects, which may be particularly difficult to quantify when the data set is inhomogeneous, and/or when the distribution of the intrinsic loudness of the signals is not known.  Moreover, in some cases there can be uncertainty in the interpretation of the signals.  Meanwhile, for systems that are not yet merging, there may be uncertainties in their future evolution.  We discuss these challenges below.


\input{table-BHNS.tex}

\subsection{Gravitational waves}

Gravitational-wave data should, at first glance, enable straightforward rate estimates.  These are direct observations of compact-object coalescences.  The data are homogeneous, with well-understood selection effects which can be modelled by injecting mock signals into the data stream and measuring how many of them will be successfully extracted by the search pipelines.  Once the surveyed time$\times$volume of the instruments is thus estimated, the volumetric merger rate is given by the number of detections in this time$\times$volume.  We might therefore expect that analysing some 70 confident BH-BH mergers from the first, second, and third advanced detector observing run in \citet{GWTC3:pop} would enable the BH-BH merger rate to be inferred to a fractional accuracy of $1/\sqrt{70} \approx 12\%$ based on Poisson statistics.

However, there are several complicating factors to this story, as a result of which the total binary black hole merger rate is still uncertain at the factor of two level.  Firstly, the detection rate is a sensitive function of mass, with more massive binaries typically providing higher-amplitude gravitational-wave signals that can be detected through a greater volume.  Therefore, inferring the total merger rate relies on simultaneously measuring the mass distribution of merging black holes, which requires many more observations \citep{GWTC3:pop}.  One way to partially overcome this uncertainty is to focus on observations in a more narrow mass band \citep{Roulet:2020,GWTC2:pop}; for example, \citet{Roulet:2020} find that the merger rate of 20--30\,$M_{\odot}$ BHs\footnote{We generally use solar masses ($M_\odot$) and solar radii ($R_\odot$) as mass and distance units.} is 1.5--5.3 \Gpcyr, while \citet{GWTC2:pop} find a BH-BH merger rate in the same mass range of $3.7^{+1.6}_{-1.1}$ or $2.6^{+1.9}_{-1.3}$ \Gpcyr depending on the assumed mass distributions.  The inferred merger rates reported in the summary tables and figures as ``PDB (ind)'', ``MS'', and ``BGP'' refer to the corresponding mass distribution models described in \citet{GWTC3:pop}: Power law + dip + break (with independent mass pairings for the two components),  Power law + peak multi-source (separate BH mass models in BH-BH and NS-BH binaries) and Binned Gaussian process (unmodelled with regularisation).

Secondly, another uncertainty arises from the possibility that the merger rate varies with redshift, for which there is already significant support in the data \citep{Fishbach:2021,GWTC3:pop}.  Allowing for the variation of the merger rate with redshift (but not yet of the mass distribution with redshift, although some theoretical models predict this) further shifts the binary black hole merger rate at redshift $z=0$.  The ``$z$-dependent'' rate reported for BH-BH mergers comes from the redshift-dependent merger rate model evaluated at $z=0$ \citep{GWTC3:pop}.

Thirdly, details of search methodology and tuning can impact the set of detections \citep[e.g.,][]{Abbott:2021-GWTC-2-1,GWTC3}, and independent searches over the publicly released LIGO-Virgo data have yielded slightly different source populations \citep[e.g.,][]{Zackay:2019,Venumadhav:2020,Nitz:2021,Nitz:2021-4OGC}.  While searches with slightly different sensitivities and different numbers of detections should still yield consistent rate estimates, some fluctuation is to be expected, especially through the impact of different inferred mass distributions as described above.

\begin{figure*}
    \centering
\includegraphics[width=1\textwidth]{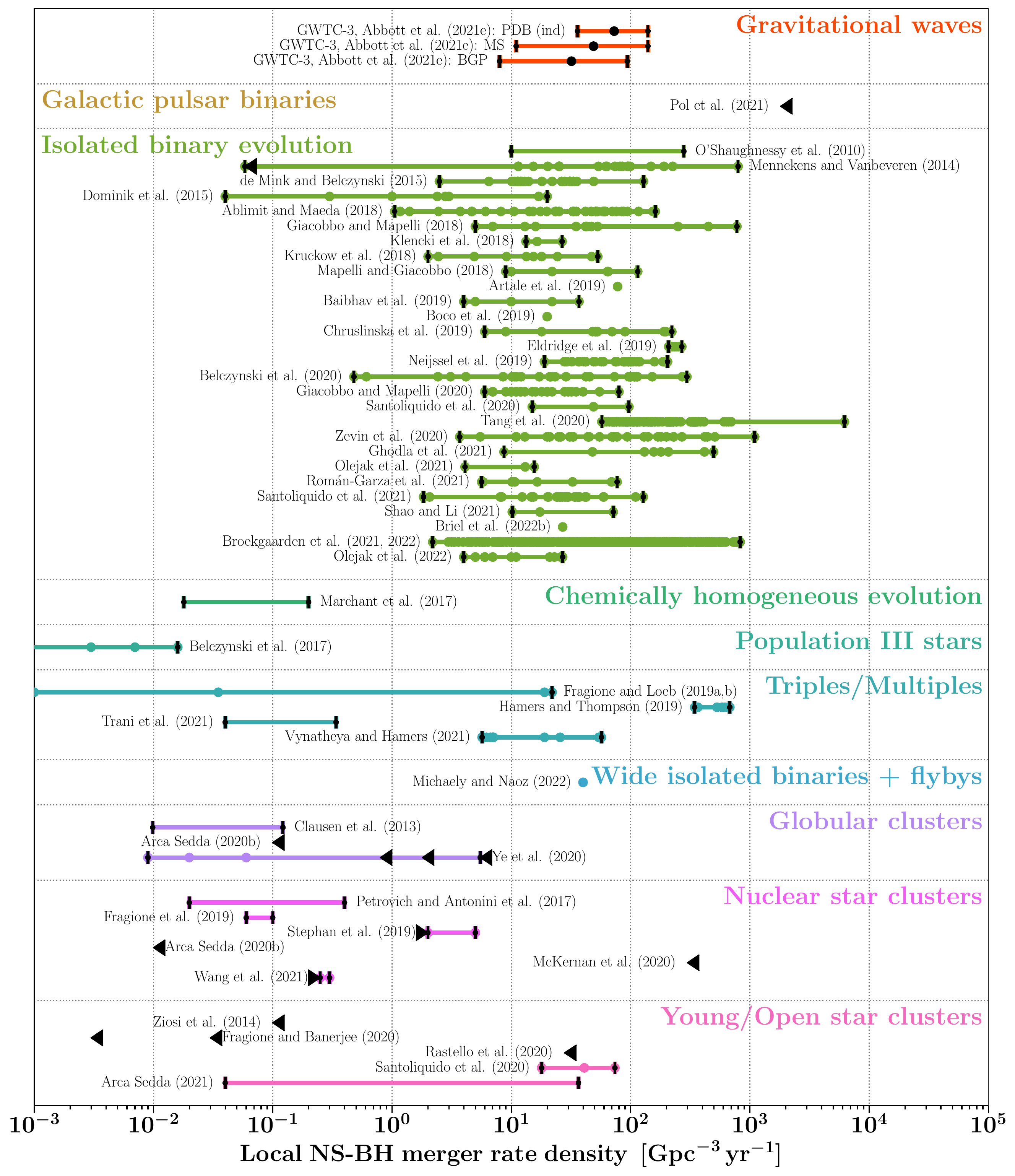} 
    \caption{NS-BH merger rates from Table~\ref{table:NSBH}.  Notation as in Figure~\ref{fig:NSNS}.
(\href{https://github.com/FloorBroekgaarden/Rates_of_Compact_Object_Coalescence/blob/main/plottingCode/Make_figures_Mandel_and_Broekgaarden_2021_COC_rates_review.ipynb}{GitHub})
    }
    \label{fig:BHNS}
\end{figure*}

Fourthly, while the chirp mass $m_1^{3/5} m_2^{3/5} (m_1+m_2)^{-1/5}$ is measured accurately for low-mass binaries through gravitational-wave signals, the mass ratio measurement is much less accurate, making it difficult to distinguish neutron stars from low-mass black holes (see, e.g., \citealt{Hannam:2013}).  A measurement of tidal deformation could point unambiguously to a neutron star rather than a black hole, but such measurements require sensitivity at higher frequencies, and only upper limits have been placed so far \citep{GW170817}.  Therefore, in the absence of an electromagnetic counterpart, it is not clear how to classify a system such as GW190814 \citep{GW190814}, which consisted of a $\approx 23\, M_\odot$ BH coalescing with a $\approx 2.6\, M_\odot$ compact object that is probably a low-mass black hole but could be a very high-mass neutron star.

Estimates of the NS-NS and NS-BH merger rates are based on a very small number of confident detections, including GW170817 and GW190425 in the former category and GW200105 and GW200115 in the latter category \citep{GW170817,GW190425,GW200105}.   Furthermore, these estimates are particularly sensitive to assumptions about the poorly constrained mass distributions in NS-NS and NS-BH binaries.


\input{table-BBH.tex}

\begin{figure*}
    \centering
\includegraphics[width=1\textwidth]{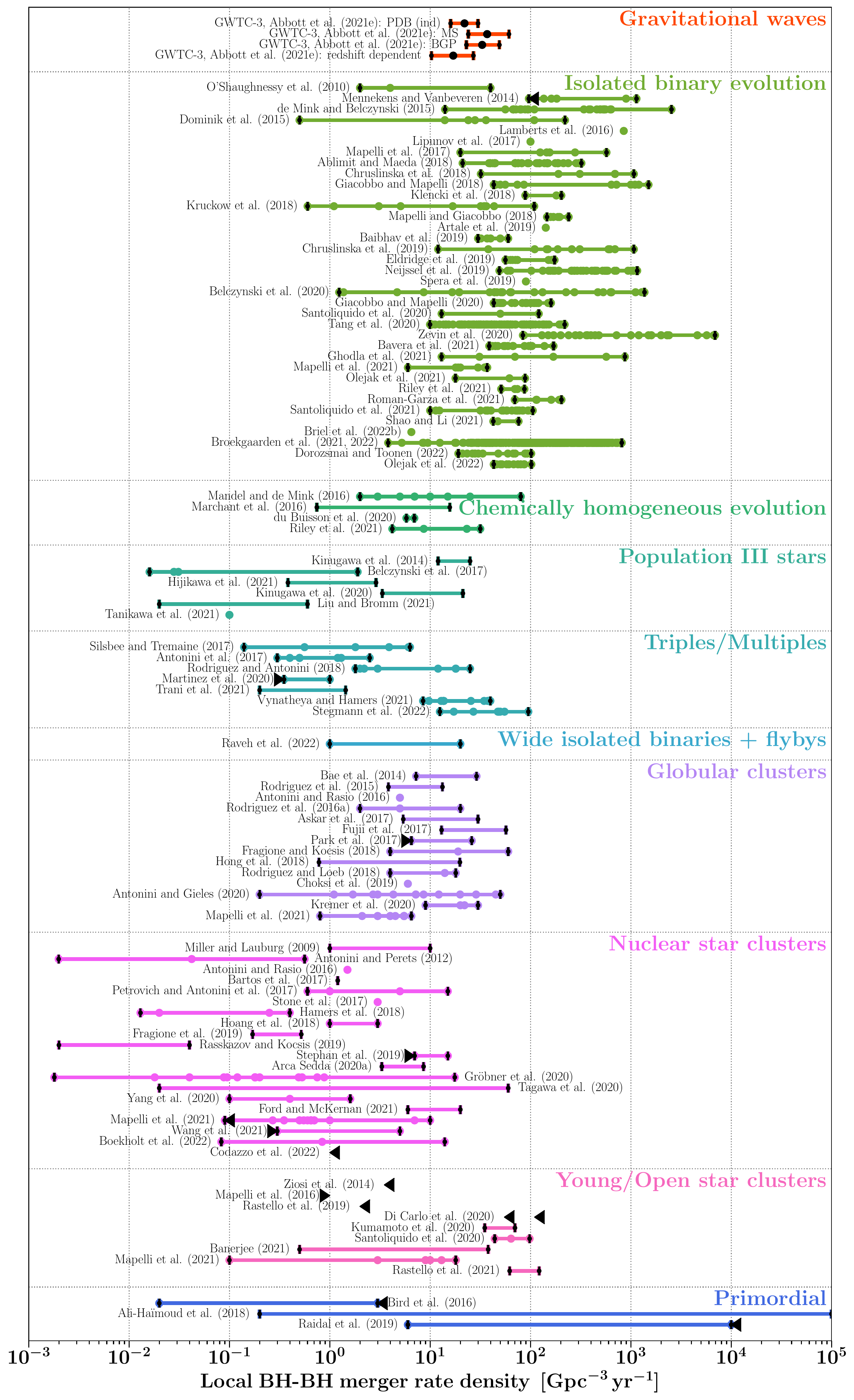} 
    \caption{BH-BH merger rates  from Table~\ref{table:BBH1}.  Notation as in Figure~\ref{fig:NSNS}. 
(\href{https://github.com/FloorBroekgaarden/Rates_of_Compact_Object_Coalescence/blob/main/plottingCode/Make_figures_Mandel_and_Broekgaarden_2021_COC_rates_review.ipynb}{GitHub})
  }
    \label{fig:BHBH}
\end{figure*}

\subsection{Short gamma-ray bursts and kilonovae}

Mergers of two neutron stars have long been expected to be accompanied by a short burst of spectrally hard gamma radiation accompanying an ultra-relativistic jet, with subsequent synchrotron radiation afterglow from the interaction of the jet with the circumstellar medium \citep{PaczynskiRhoads:1993,Rosswog:2002,nakar07}.  Meanwhile, the formation and subsequent radioactive decay of neutron-rich elements through r-process nucleosynthesis was conjectured to power an optical kilonova \citep{LiPaczynski:1998,Rosswog:1999,Metzger:2010,Kasen:2013}.  Both expectations were spectacularly confirmed with the  detection of a short gamma-ray burst \citep{GW170817:GRB}, followed by the observation of a kilonova \citep{GW170817:MMA, Kasen:2017, Metzger:2019kilonova}, and finally a broad-spectrum (radio through optical to X-ray) afterglow \citep{Mooley:2017,Lyman:2018,Troja:2017} following the NS-NS merger GW17017.

One concern with using short gamma-ray bursts (SGRBs) to infer binary neutron star coalescence rates is the uncertainty in a one-to-one association between these events.  On the one hand, some SGRBs and perhaps even kilonovae could be associated with NS-BH mergers \citep{Troja:2008,Berger:2014,Li:2017,Gompertz:2020}.  However, this is likely only possible for those events when the neutron star is tidally disrupted at a sufficient distance from the black hole, requiring the black hole to be low in mass and/or rapidly rotating \citep{Foucart:2018}.  Therefore, the contribution of NS-BH mergers to SGRBs is likely low \citep[e.g.,][]{Drozda:2020}.   On the other hand, given the challenges in reproducing ultra-relativistic jet formation in numerical merger simulations \citep{Mosta:2020}, it cannot be determined with confidence whether all neutron star mergers should be accompanied by SGRBs.

Nevertheless, SGRBs have been used for more than a decade as proxies for the NS-NS merger rate, with \citet{Nakar:2006} estimating the merger rate as at least 10 Gpc$^{-3}$ yr$^{-1}$, but possibly several orders of magnitude higher.  The key challenge is the uncertainty in the angular size of the jet: the smaller the jet, the smaller fraction of SGRBs beamed toward Earth, and hence the larger the intrinsic rate based on the observed rate of SGRBs.  \citet{Fong:2012} used evidence for jet breaks to constrain the angular size of the jet, allowing them to infer an NS-NS merger rate.  However, the evidence from GW170817 firmly indicated that the jet has spatial structure, rather than a simple top-hat profile \citep{Troja:2018,Resmi:2018,Lamb:2018obs}.  This requires a re-evaluation of earlier models, an ongoing effort \citep{Paul:2018}; for example, \citet[][see their figure 3]{MarguttiChornock:2020} estimate the rate of GW170817A-like SGRBs as $\gtrsim 300$ \Gpcyr given the observed off-axis angle  under the assumption that other SGRBs would have similar jet profiles.  

The first kilonova observation was made in 2013 \citep{Tanvir:2013}.  While more candidates have been observed in the last few years, they are still difficult to unambiguously distinguish from other transients. 
At present, their luminosity function is not yet sufficiently well known to infer an accurate rate.  However, the absence of transients with a kilonova signal similar to that accompanying GW170187 in data from the Palomar Transient Factor allowed \citet{Kasliwal:2017} to place a conservative 3-$\sigma$ upper limit of 800 Gpc$^{-3}$ yr$^{-1}$. \citet[][see their Fig.~9]{Andreoni:2021} compile a list of optical kilonova surveys with upper limits between 900 and 99000 \Gpcyr.

\subsection{Pulsars in Galactic compact object binaries}

Observations of double compact object binaries prior to merger are the next best thing to direct merger observations for measuring merger rates.  Provided that the masses, period or separation, and eccentricity of the binary are known, the time to merger through the emission of gravitational waves can be computed  \citep{Peters:1964}.  Assuming a steady-state configuration, which is a reasonable assumption for our Galaxy with its slowly evolving metallicity and star formation rate, an observed population of pre-merger compact object binaries yields an estimate of the coalescence rate.

So far, NS-NS binaries are the only double compact objects for which direct observations are available.  Nearly 20 such binaries have been observed in the Galaxy through radio pulsar pulsar observations  \citep[see, e.g.,][for recent compilations]{Tauris:2017,Farrow:2019}.  Of course, these are expected to represent a relatively small fraction of all Galactic double neutron stars, with most remaining unobserved because neither neutron star is a pulsar or any pulsars are too dim or beamed away from the Earth.  Moreover, pulsars in the tightest binaries are rare because of their rapid inspiral and may be challenging to detect because the pulsations are Doppler shifted by orbital modulation.  Therefore, the key challenge in inferring the merger rate of close double neutron stars in the Galaxy from these observations lies in accounting for selection effects, particularly the uncertain luminosity function and beaming fraction of pulsars.

Fortunately, the rapidly growing amount of pulsar data since the pioneering efforts of \citet{Phinney:1991ei} and \citet{Narayan:1991} to infer NS-NS merger rates from binary pulsar observations enables increasingly accurate treatments of these selection effects.  Much of the recent literature follows the statistical framework of \citet{Kim:2003kkl}, with the addition of varying models of the luminosity distribution \citep{Kalogera:2004tn}, more sophisticated models of the pulsar beaming fraction which could vary with pulsar age \citep{OShaughnessyKim:2009}, an attempt to account for the observation of both pulsars in the double pulsar J0737-3039 \citep{Kim:2015} and the addition of new systems (with the relatively small number of known Galactic binaries, a single new short-lived double neutron star could impact the rate estimate at the factor of two level, as shown by \citealt{Kim:2006}).  We report the rates from a few recent papers to indicate the evolving estimates of the double neutron star merger rate \citep[see][for a compilation including older results]{ratesdoc}.

There are several complicating factors in extracting coalescence rates from the observed Galactic double neutron stars.  B2127+11C and possibly J1807-2500 are known double neutron stars in Galactic globular clusters; these are generally excluded from rate estimates because they are believed to be rare dynamically formed systems \citep[][but see \citealt{AndrewsMandel:2019}]{Ye:2019}.  Another concern is that the detectability of a Galactic double neutron star may correlate with the evolutionary channel and with the system properties, e.g., through the amount of recycling the pulsar experiences.  This could lead some sub-populations of pulsars -- perhaps the more massive ones like GW190425 \citep{GW190425,RomeroShaw:2020,VignaGomez:2021DNS} -- to be missed in Galactic surveys.  Moreover,  extrapolation from the Milky Way merging double neutron star rate to a volumetric rate depends on the accuracy of estimating the local star formation rate \citep{pacheco:2005} and on the sensitivity of the yield of merging double neutron stars to metallicity.  

No Galactic radio pulsars in NS-BH binaries have been detected so far.  However, the lack of such observations only weakly constrains the NS-BH merger rate from above.  If the heavier BH forms first, which appears intuitive given the shorter lifetimes of more massive stars, the NS will not be recycled by accretion, which limits the time during which it can be observed as a radio pulsar \citep{Debatri:2021}.  Mass transfer can invert the mass ratio, allowing the NS to form first from the originally more massive star; however, the formation rate of such systems is likely quite low \citep{Pfahl:2005}.  

\subsection{High-mass X-ray binaries}

As we will discuss in the next section, most compact object binaries formed through isolated binary evolution are expected to go through a high-mass X-ray binary stage, in which the first companion to form a black hole or neutron star accretes from the wind of the stellar companion.  This phase may directly precede the formation of two compact objects, and can therefore be used to constrain compact binary coalescence rates.  For example, \citet{Bulik:2008} used the extragalactic high-mass BH X-ray binaries IC10 X-1 and NGC300 X-1 to estimate the merger rate of BH-BH binaries at $360^{+500}_{-260}$ Gpc$^{-3}$ yr$^{-1}$.  

However, such estimates suffer from significant uncertainties.  Firstly, there are often large systematic errors in the masses and separations of the observed systems; in particular, the masses of IC10 X-1 and NGC300 X-1 may be much lower than in the analysis of \citet{Bulik:2008} because orbital velocities are very challenging to measure in the presence of optically thick Wolf-Rayet winds \citep{Laycock:2015}.  Even well-studied high-mass X-ray binaries such as Cygnus X-1 \citep{Orosz:2011} may require significant corrections to their masses \citep{MillerJones:2021}, while the uncertain present properties of Cygnus X-3 translate into uncertainty about its future fate \citep{CygnusX3:2012}.  Secondly, the observed systems typically come from a heterogeneous set of observations, complicating the determination of selection effects.  Thirdly, the future evolution of such systems is sensitive to assumptions about the dynamical stability of any future mass transfer episodes \citep{vdH:2017}, the amount of mass loss in winds (cf.~\citealt{Belczynski:2011CygX1} and \citealt{Neijssel:2020CygX1} for the case of Cygnus X-1), and the remnant mass and natal kick of the remaining supernova.  

\citet{Inoue:2016} and \citet{FinkeRazzaque:2017} use observations of ultra-luminous X-ray sources (ULXs) with X-ray luminosities exceeding $10^{39}\,\rm{erg}\,\rm{s}^{-1}$ to estimate BH-BH merger rates of $\lesssim 100$\Gpcyr  and  $15$--$400$\Gpcyr, respectively.  These estimates, however, are highly uncertain. They rely on the assumption that all ULXs are powered by accretion onto a BH, which is contradicted by observations that at least some ULXs (perhaps all with confidently identified compact objects?) are powered by accretion onto an NS \citep{Bachetti:2014}.  They also make simplifying assumptions about the future evolutionary history of ULXs.

\subsection{Other observational constraints}

\citet{ratesdoc} suggested that a strict upper limit of $<5 \times 10^{4}$ mergers Gpc$^{-3}$ yr$^{-1}$ could be placed on the NS-NS merger rate by assuming that at least one companion must have been formed through a type Ib/Ic supernova (taking an optimistic rate estimate from \citealt{Cappellaro:1999}).  However, this upper limit is likely a very significant over-estimate, as most type Ib/Ic supernovae happen in other systems than merging binary neutron stars (perhaps even in systems that had already been disrupted, \citealt{Hirai:2020}).  At the same time, the type Ib/Ic supernova rate cannot even be considered a strict upper limit, because many merging neutron star binaries could form through a combination of electron-capture supernovae and ultra-stripped supernovae rather than classical core-collapse supernovae \citep{Podsiadlowski:2004,DallOsso:2014,Tauris:2015,Tauris:2017,VignaGomez:2018}, and some compact remnants may form in the absence of any explosions. 

Signatures of chemical enrichment associated with binary neutron star mergers, particularly the evolution of r-process nucleosynthesis elements, could serve as a useful constraint on the binary neutron star merger rate \citep{Mennekens:2014,Hotokezaka:2015,Beniamini:2016rp}.  However, \citet{Siegel:2019,Kobayashi:2020} propose that magneto-rotational supernovae could instead be responsible for r-process enrichment.

Compact-object mergers or the subsequent evolution of compact-object merger products have been proposed as a source for some fast radio bursts \citep{Totani:2013,RaviLasky:2014}; however, there are likely multiple progenitor channels for fast radio bursts \citep{Pleunis:2021}, so these cannot presently be confidently used for merger rate estimates.


\section{Modelled rates}
\label{sec:theory}

In this section, we summarise some of the ongoing efforts to model the rates of compact object mergers formed through a variety of evolutionary channels.  These channels offer alternative pathways to overcoming a key challenge in gravitational-wave astronomy.  Gravitational-wave emission is only efficient in close binaries: in order for a circular binary with equal-mass components of mass $M$ to merge within a time $T$ exclusively through radiation reaction from gravitational-wave emission, the separation $a$ must be smaller than 
\begin{equation}
a \lesssim 3.7\, R_{\odot}\ \left(\frac{M}{M_{\odot}}\right)^{3/4} \left(\frac{T}{14\ \mathrm{Gyr}}\right)^{1/4}.
\end{equation}
Thus, to merge within the current age of the Universe, 14 Gyr, two $1.4\, M_{\odot}$ neutron stars must be fewer than 5 solar radii apart, while even two $35\, M_{\odot}$ black holes must be within a quarter of an astronomical unit.  

On the other hand, massive stars typically expand to several au (hundreds to thousands of solar radii; 1 au $\approx$ 215 R$_\odot$) in size during their evolution.  Therefore, the key challenge to forming a merging compact-object binary lies in fitting a peg into a hole that is one or two orders of magnitude smaller than the size of the peg.  The proposed solutions generally take the form of either relying on stellar evolution and mass transfer to overcome the problem in an isolated binary, or circumventing the problem altogether by relying on dynamics to form binaries with a sufficiently small separation that gravitational-wave emission can drive them to merge.  Here we sketch out the key steps in the proposed formation channels, undertake back-of-the-envelope estimates of the rates, following \citet{MandelFarmer:2018}, and describe some of the key uncertainties leading to the typically broad range of model predictions.

One particular source of uncertainty that is common to all models described below is the uncertain rate and metallicity of star formation across cosmic history.  Even binaries merging in the relatively local Universe (detections through the third LIGO/Virgo observing run had moderate redshifts $z \lesssim 1$, \citealt{GWTC3}) could have formed at high redshift \citep{Belczynski:2016}.  Therefore, the rate of star formation at higher redshift influences local merger rates.  Moreover, the yield of coalescing compact objects per unit star forming mass depends strongly on metallicity \citep{Dominik:2013,Neijssel:2019}. This is particularly true for black holes, whose massive progenitors can lose large fractions of their mass in winds in high-metallicity environments, widening the binaries and reducing merger rates.  Therefore, the local merger rate is sensitive to the poorly known metallicity-specific star formation rate at high redshifts \citep{Belczynski:2010,Mapelli:2017,Klencki:2018,Chruslinska:2019,Neijssel:2019,ChruslinskaNelemans:2019,Tang:2020,Briel:2021,Broekgaarden:2021,Broekgaarden:2022,Santoliquido:2021,vanSon:2021}.
 
\subsection{Isolated binary evolution modelling (population synthesis)}

Almost all stars massive enough to form neutron stars and black holes are born in binaries or higher-multiplicity systems \citep{Sana:2012,MoeDiStefano:2017}.  We begin by considering compact-object coalescences in isolated binaries, namely those which do not appreciably interact with their environment.  This channel has been studied since the 1970s.  Some of the ground-breaking works particularly relevant to the formation of coalescing compact-object binaries include \citet{Tutukov:1973, vdHDeLoore:1973,DeLoore:1975,FlanneryvdH:1975,TutukovYungelson:1993,Lipunov:1997,BetheBrown:1998,Dewi:2006,Nelemans:2001,VossTauris:2003,Kalogera:2007,OShaughnessy:2008} (see \citealt{MandelFarmer:2018} for a brief summary of the early history).

\subsubsection{Isolated binary evolution with mass transfer}

In this channel, two massive stars are born in a wide binary, giving them sufficient space to evolve.  The binary shrinks as a consequence of mass transfer just when the stars themselves shrink in radius, allowing gravitational waves to take over.  

We begin by sketching out a very simplified version of binary evolution in this model.  The primary -- the initially more massive star -- completes core hydrogen fusion first.  At the end of the main sequence, its helium-rich core contracts and the hydrogen-rich envelope expands.  As the star fills the so-called Roche lobe, tidal gravity from the companion overcomes the self-gravity of the primary, and mass transfer onto the secondary commences.  Eventually, the primary leaves behind a naked helium star, which continues nuclear fusion, losing mass through winds, until it collapses into a compact object.  The mass loss and/or the natal kick from asymmetric mass ejection accompanying the supernova may unbind the binary.  If it does not, the binary continues its evolution until the secondary expands and commences mass transfer onto the now compact-object primary.  Because of the mass gain of the secondary during the first episode of mass transfer and the mass loss of the primary due to mass transfer, winds, and the supernova, the secondary may be significantly more massive than the primary at this stage.  Consequently, mass transfer could harden the binary \citep{vdH:2017}, and could even become dynamically unstable, leading to the formation of a common envelope \citep{Paczynski:1976}.  Once the common envelope is not co-rotating with the binary, drag forces dissipate orbital energy, which may allow for common-envelope ejection after the binary hardens by two or three orders of magnitude \citep{Ivanova:2013}.  At this stage, the secondary has lost its envelope to become a naked helium star, so it can fit into the tight binary.  (It is also possible that both stars are evolved giants during the first mass transfer interaction, leading to the simultaneous ejection of both envelopes during a single common-envelope event, as described by \citealt{Dewi:2006}.)  After more mass loss through winds the secondary may also collapse into a compact object.  Assuming the second supernova leaves behind a bound, compact binary, this binary will eventually merge by emitting gravitational waves.  

We can crudely estimate the expected merger rate through this channel with a Drake-like equation describing the fraction of all binaries that go on to make merging double compact objects of a particular type \citep{MandelFarmer:2018}, with chosen values corresponding to binary black hole formation:
\begin{align}
f_\textrm{DCO} &= f_\textrm{primary} \times f_\textrm{secondary} \times f_\textrm{init sep} \times f_\textrm{survive SN1} \times f_\textrm{CE} \times f_\textrm{survive SN2} \nonumber \\  & \times f_\textrm{merge} 
  \sim 0.001 \times 0.5 \times 0.5 \times 1 \times 0.1 \times 1 \times 0.2 = 5 \times 10^{-6}. \label{DrakeIsolated}
\end{align}
Assuming a local star formation rate of $\sim 10^7\, M_{\odot}$ Gpc$^{-3}$ yr$^{-1}$ \citep{MadauDickinson:2014} with an average mass of $\sim M_{\odot}$, a yield of $f_\textrm{DCO} = 5 \times 10^{-6}$ would lead to a compact-object merger rate of $\sim 50$ Gpc$^{-3}$ yr$^{-1}$.

Here we describe the terms in this equation, which allows us to highlight the broad range of physics involved and to comment on the key uncertainties.  In practice, most modern models rely on some form of population synthesis, where large numbers of stellar binaries are evolved in order to compute the yield, frequently by using simplified recipes to reduce computational cost \citep{PostnovYungelson:2014,Han:2020}.

Stars with initial mass at zero age on the main sequence of roughly $\gtrsim 20\, M_{\odot}$ can form a BH, so the fraction of all stars drawn from the initial mass function that will have mass in this range is $f_\textrm{primary} \approx 0.001$.  Meanwhile, for forming an NS, the zero-age main sequence mass should be roughly between 8 and $20\, M_{\odot}$, so $f_\textrm{primary} \approx 0.002$.  The mass ratio between secondary and primary is roughly uniformly distributed \citep{Sana:2012}, so if the primary falls in the mass range of interest, roughly half the secondaries will do so, i.e., $f_\textrm{secondary} \approx 0.5$.  The fraction of binaries with initial separations in the range to allow for the first episode of mass transfer from evolved donors is quite large,  $f_\textrm{init sep} \approx 0.5$, given the significant amount of expansion during this phase and the uniform in the logarithm distribution of initial separations \citep{Opik:1924}.  

The probabilities of the binary surviving the first and second supernovae without being disrupted, $f_\textrm{survive SN1}$ and $f_\textrm{survive SN2}$, are both close to 1 for the heavier black holes that may form through complete collapse with negligible natal kicks \citep[][but see \citealt{Repetto:2012,Mandel:2015kicks,Atri:2019}]{Mirabel:2016}.  However, the high typical natal kicks of neutron stars, of order 300 km s$^{-1}$ \citep{Hobbs:2005,IgoshevVerbunt:2017}, can disrupt $\gtrsim 95\%$ of wide binaries in the first supernova \citep{VignaGomez:2018,Renzo:2019}.  Reduced kicks in electron-capture supernova \citep{Podsiadlowski:2004,Gessner:2018} may protect some of the wide binaries from unbinding, increasing $f_\textrm{survive SN1}$.  By the time of the second supernova, the binary will have gone through a common envelope in this channel, hardening it and making disruption less likely; moreover, low-mass helium star progenitors of neutron stars may re-expand after the helium main sequence, engage in another mass transfer episode (case BB mass transfer) and ultimately explode in very weak ultra-stripped supernovae with greatly reduced kicks \citep[][but see \citealt{MandelMueller:2020}]{Tauris:2015}.  Therefore, it is plausible that $f_\textrm{survive SN2} \approx 1$, even for double neutron stars, although observed eccentricities of $\sim 0.6$ for a subset of Galactic double neutron stars, including the Hulse--Taylor binary, could point to a subpopulation of systems that avoid ultra-stripping and may be disrupted by the second supernova.

The common-envelope phase is perhaps the least certain one in massive binary evolution  \citep{Dominik:2012}.  There are questions about both the onset of the common-envelope phase (do the mass ratio and the evolutionary state of the donor and accretor make mass transfer dynamically unstable?) and about the ultimate outcome of the evolution (is sufficient energy deposited in the envelope to eject it, or does this episode end in merger?).  Typical population synthesis models based on current, very simplified recipes for both the stability mass transfer and the survivability of the common-envelope phase yield $f_\textrm{CE} \approx 0.1$.

The fraction of compact-object binaries that ultimately coalesce in the age of the Universe $f_\textrm{merge}$ is set by the distribution of masses and separations.  If the post-common-envelope binary separation is distributed uniformly in the logarithm, the delay time $\tau$ between formation and merger, which scales with the fourth power of the separation, will also follow a log-uniform distribution, i.e., $p(\tau) \propto 1/\tau$.  In that case, the logarithmic distribution allows for typical $f_\textrm{merge} \approx 0.2$; however, as mentioned above, this can be significantly reduced, e.g., if winds increase the BH binary separation at high metallicity.

We now describe the key uncertainties in these estimates.  Different treatments of some of the physics summarised below are responsible for the broad range of predictions reported in the executive summary.

The first significant uncertainty lies in the distribution of initial conditions: the masses of the two companions, their separations, and the binary eccentricity \citep{Sana:2012,MoeDiStefano:2017}.  These alone can impact the merger rate at the level of a factor of $\sim 6$ due to uncertainty in the initial mass function \citep{Kroupa:2002} and by up to a factor of 2--3 due to uncertainty in the other parameters \citep{deMinkBelczynski:2015,Klencki:2018}. 

The treatment of mass transfer represents a key source of uncertainty in isolated binary evolution \citep{Belczynski:2021}.  Is the mass transfer dynamically stable?  Stability thresholds can be formulated in terms of the mass ratio \citep{Claeys:2014} or the relative radial response of the star and its Roche lobe to mass transfer \citep{Soberman:1997,Ge:2015}.  The depth of the convective envelope of the donor may play a critical role \citep{Klencki:2020convective}.  
If the mass transfer is dynamically stable, how conservative is it, i.e., what fraction of mass lost by the donor does the companion accrete \citep[e.g.,][]{KippenhahnMeyerHofmeister:1977}, including the case of accretion onto compact objects \citep[e.g.,][]{vanSon:2020}?  And how much angular momentum is carried away by any mass that leaves the system, changing the orbital separation of the binary \citep[e.g.,][]{Vinciguerra:2020}?  How does mass transfer in eccentric binaries affect their orbit \citep[e.g.,.][]{Sepinsky:2010,Dosopoulou:2016}?
If the mass transfer is dynamically unstable, leading to the formation of a common envelope, can the envelope be ejected, and what sets the orbital separation after the envelope ejection?  Simple energy scalings are typically used to treat the common-envelope phase of binary evolution \citep{Webbink:1984,deKool:1990}, partly because, despite decades of modelling common envelopes \citep{Ivanova:2013}, the first incomplete models of this phase for massive stars are only just appearing \citep{Ricker:2019,LawSmith:2020,Lau:2021}.  Last but not least, how does the history of mass transfer impact the final outcomes of stellar evolution, including supernova explosions \citep[e.g.,][]{Brown:2001,Schneider:2020}? 

A range of uncertainties in single stellar evolution play a key role.  How much massive stars expand at various stages in their lives determines the onset and outcomes of mass transfer \citep[e.g.,][]{Laplace:2020}.  Stellar winds affect the final masses and separations \citep{Vink:2017}.  The efficiency of tidal coupling determines orbital circularisation \citep{Zahn:1977,Hut:1981,VickLai:2018}. The amount of mass lost in supernovae and the natal kicks that supernova remnants receive impact binary survival during the stellar explosions \citep{Sipior:2002,Fryer:2012,Mueller:2020}.  Even relatively rare supernova variants such as electron-capture supernovae and (pulsational) pair-instability supernovae are important for coalescing compact object formation \citep[e.g.,][]{Belczynski:2018,Stevenson:2019,Farmer:2020}.

\subsubsection{Isolated binaries: chemically homogeneous evolution and population III stars}

One way to avoid the problem of fitting very extended stars into a binary tight enough to merge through gravitational waves is to prevent the stars from expanding in the first place.  Here we consider two magic wands that might prevent expansion: efficient mixing or initially metal-free composition.

Suppose the stars are sufficiently massive, their metallicity is low enough to suppress strong winds, and they are close enough to be tidally locked in an orbit that enforces rotation at a significant fraction of the break-up velocity.  Then the formation of temperature gradients between the poles and equator of the star may lead to large-scale circulation and efficient mixing \citep{Eddington:1925,Sweet:1950,EndalSofia:1978}.  The mixing of helium out of the core into the envelope and fresh hydrogen from the envelope into the core allows the star to evolve chemically homogeneously, fusing almost all of the hydrogen into helium on the main sequence \citep{Heger:2000,MaederMeynet:2000,Yoon:2006}.  After the end of hydrogen fusion, the entire star contracts into a naked helium star.  Such systems avoid mass transfer after the main sequence, and may form heavy, ultimately coalescing black hole binaries in situ \citep{MandelDeMink:2016,Marchant:2016,deMinkMandel:2016}.  The high mass required for chemically homogeneous evolution makes this channel relevant only for the formation of merging black holes and, perhaps, rare NS-BH binaries \citep{Marchant:2017}.

We can write a similar Drake's equation to Eq.~(\ref{DrakeIsolated}) for the chemically homogeneous evolution channel.  Following \citet{MandelDeMink:2016}, the fraction of binaries that form coalescing binary black holes through this channel is
\begin{align}
f_\textrm{BH-BH} &= f_\textrm{primary} \times f_\textrm{secondary} \times f_\textrm{init sep} \times f_Z \times f_\textrm{merge} \nonumber \\
 & \sim 0.0002 \times 0.5 \times 0.1 \times 0.1 \times 1 = 10^{-6}.
\end{align}
Only the most massive stars with initial masses above $\sim 50\, M_{\odot}$ are likely to experience sufficient mixing to evolve chemically homogeneously, lowering $f_\textrm{primary}$ to $\approx 2 \times 10^{-4}$.  A flat mass ratio distribution would again yield $f_\textrm{secondary} \approx 0.5$.  The range of initial separations allowing for chemically homogeneous evolution is at most a factor of two: stars that are initially too far apart either avoid tidal locking or do not rotate sufficiently rapidly, while stars that are too close overflow the L2 Lagrange point and promptly merge.  If initial separations cover some 5 orders of magnitude and a uniform in the logarithm distribution of separations is assumed, up to $f_\textrm{init sep} \approx 0.1$ of binaries with the right mass range could still experience chemically homogeneous evolution.  However, this also requires a sufficiently low metallicity, in part to avoid binary widening through excessive mass loss in winds, necessitating a factor describing the fraction of star formation at metallicities of interest, estimated here as $f_Z \approx 0.1$.  On the other hand, if a binary evolves chemically homogeneously and forms black holes in situ, it is likely to be sufficiently compact to merge within the age of the Universe through gravitational-wave emission, so $f_\textrm{merge} \approx 1$.  The overall yield estimate of $f_\textrm{BH-BH} \approx 10^{-6}$ would suggest a binary black hole coalescence rate of $\sim 10$ Gpc$^{-3}$ yr$^{-1}$ for this channel.

The chemically homogeneous evolution channel as described above does not involve mass transfer after the main sequence and hence avoids some of uncertainties discussed in the previous subsection.  On the other hand, the very possibility of efficient chemical mixing is uncertain.  The observational evidence ranges from tentative to contradictory \citep{Almeida:2015,MandelDeMink:2016,AbdulMasih:2021} and theoretical models of mixing differ in the predicted thresholds on its efficiency.  Moreover, much of the parameter space in this channel likely requires over-contact binaries \citep{Marchant:2016,duBoisson:2020,Riley:2020}, which bring extra modelling challenges and uncertainties.   On the other hand, a star could be spun up to rapid rotation and chemically homogeneous evolution by mass accretion from the companion \citep{Cantiello:2007}.  This mode of quasi chemically homogeneous evolution could also make a significant contribution to high-mass BH-BH mergers originating in low-metallicity environments \citep{EldridgeStanway:2016}.

First-generation metal-free stars in the early Universe, known as population III stars, may represent an alternative way to avoid radial expansion after the main sequence.  The absence of primordial carbon, nitrogen and oxygen in these stars prevents the CNO cycle from operating, so that hydrogen fusion can proceed only through the pp-chain reaction.  While the core becomes hot enough for helium produced in the core to form carbon through the triple-alpha reaction, eventually allowing CNO cycle hydrogen fusion to take over, this may not happen to the same extent in the hydrogen shell around the helium core after core hydrogen exhaustion.  Therefore, population III stars may experience reduced radial expansion relative to more metal-rich stars \citep{Marigo:2001} and can avoid mass transfer and many of the complications of the classical isolated binary evolution channel.  Moreover, this means that the hydrogen envelope of such stars remains tightly bound, and is unlikely to be ejected during the stellar collapse as would be the case for population I and II supergiants \citep{Nadyozhin:80,Lovegrove:2013,Fernandez:2018} allowing the star to retain most of its mass \citep{Kinugawa:2021}, akin to chemically homogeneously evolving stars.  This mass retention is further aided by the drastically reduced wind-driven mass loss at zero metallicity.  Finally, the lack of metals has been conjectured to suppress the fragmentation of a giant molecular cloud during star formation, increasing the typical mass of population III stars and making these prime progenitor candidates for binary black hole coalescences \citep{Belczynski:2004popIII,Kinugawa:2014,Inayoshi:2016,Inayoshi:2017}.

The lack of observational evidence and challenges in theoretical models put many of these assumptions in doubt.  There are conflicting models for the initial properties of population III binaries \citep{Hirano:2014,Stacy:2016} which could drastically reduce the predicted merger rate \citep{Hartwig:2016, Belczynski:2017popIII}, while the possibility of significant stellar expansion could remove this altogether as an independent channel.

\subsection{Dynamical formation}

Dynamical formation broadly encompasses compact object mergers in which gravitational interactions with other stars played a significant role in bringing the two merging objects closer.  Dynamical formation can be broadly divided into several categories based on environment.  Historically, the focus has been primarily on young star clusters and globular clusters, where dynamical formation has been explored by \citet{Sigurdsson:1993,Kulkarni:1993,PortegiesZwart:2000,OLeary:2006,Banerjee:2010,Downing:2011,Morscher:2015} through a mix of numerical gravitational experiments and semi-analytical estimates.  We illustrate the expectations for globular cluster dynamical formation with another Drake's equation before discussing other dynamical formation environments: galactic nuclei and hierarchical triple systems.

\subsubsection{Young star clusters and globular clusters}

The compact objects in the cluster may or may not have started out in the same binary.  In the classical version of dynamical formation in a dense stellar environment the black holes, being heavier than other cluster objects, sink toward the centre of the cluster through mass segregation \citep{Spitzer:1969,BinneyTremaine}.  Once there, they readily form binaries or substitute into existing binaries through three-body interactions: in a binary-single interaction, the lightest object is most likely to be ejected, while the two heavier objects remain behind in a binary \citep{SigurdssonPhinney:1993,GinatPerets:2020}.  If the binary is hard -- i.e., if its orbital velocity is larger than the velocity dispersion for the case of comparable component and interloper masses -- subsequent interactions will further harden the binary as the energy of the binary is shared with the interlopers.  Eventually, the binary may harden enough for gravitational waves to take over, provided the interaction rate is sufficiently high and the binary is not prematurely ejected from the cluster.   This suggests that the number of BH-BH mergers per globular cluster is:

\begin{align}
N_\textrm{BH-BH,\ cluster} &= N_\textrm{stars} \times f_\textrm{BH} \times f_\textrm{retain} \times f_\textrm{merge} \nonumber \\
 & \sim 10^6 \times 0.001 \times 0.2 \times 1 = 200.
\end{align}

Here, we focus on BH-BHs, because neutron stars, being lighter, are preferentially ejected during interactions, reducing their formation rate \citep{Ye:2019}.  We consider the heaviest globular clusters with $N_\textrm{stars} \approx 10^6$ stars.  We optimistically assume that all $\gtrsim 20\,\mathrm{M}_\odot$ stars will form a BH, ignoring the likelihood that natal kicks will eject some of the BHs from the cluster, whose escape velocity is only $\sim 50\,$km s$^{-1}$.   This sets $f_\textrm{BH} \approx 0.001$.  In order for a binary to merge through the emission of gravitational-waves within 10\,Gyr, its orbital velocity must be $\sim 500\,$km s$^{-1}$ \citep{Peters:1964}.  This is almost two orders of magnitude larger than the typical velocity dispersion of 10\,km s$^{-1}$.  Tens of strong binary-single scattering interactions, mostly with other black holes, will be necessary to harden one binary by two orders of magnitude \citep{Quinlan:1996}.  Once the binary becomes sufficiently hard, the lightest of the three interacting objects will typically be ejected from the cluster, depleting the fraction of retained black holes by a factor of a few \citep{HeggieHut:2003}.  We therefore set the fraction of black holes retained for possible mergers to $f_\textrm{retain} \approx 0.2$.  Finally, we consider a cluster sufficiently dense (central BH density above $\sim 10^5$ pc$^{-3}$) that most of the BHs that avoid being ejected do succeed in merging within 10\,Gyr, so that $f_\textrm{merge} = 1$.  Even with these optimistic assumptions, and allowing for a constant merger rate over the 10\,Gyr lifetime of globular clusters, whose space density is $\sim 1\,$Mpc$^{-3}$, we estimate a maximum merger rate of $200 \times 10^9\ \mathrm{Gpc}^{-3} \times 10^{-10}\ \mathrm{yr}^{-1} = 20$\Gpcyr. 

This very crude estimate helps highlight some of the issues that may impact the event rates.  What is the fraction of compact objects retained in clusters despite natal kicks?  What were the initial properties (mass distribution, binary fraction, density, velocity dispersions) of globular clusters, before the $\gtrsim 10\,$Gyr of evolution typical for the locally observed clusters \citep{FragioneKocsis:2018}?  What fraction of merger products avoid being ejected by gravitational-wave recoil kicks and can be reused for subsequent hierarchical mergers \citep{Rodriguez:2018,GerosaFishbach:2021}?  What role does ongoing stellar evolution, including BH growth through tidal disruption of stars and gas accretion \citep{Vesperini:2010}, play in dynamical formation?  While multiple approaches to modelling cluster gravitational dynamics (e.g., N-body and Monte Carlo models) are expected to yield consistent results, the range of predictions highlighted in the executive summary tables and figures illustrates the impact of different approaches to answering the questions posed in this paragraph.

\subsubsection{Nuclear star clusters}

Dynamical formation has also been explored in the context of nuclear clusters in the centres of galaxies.  \citet{MillerLauburg:2008} and \citet{AntoniniRasio:2016} pointed out that nuclear clusters in the absence of a massive black hole, which are characterised by deeper potentials and hence higher escape velocities than globular clusters, could be promising environments for BH-BH dynamical formation.   \citet{OLeary:2008,Tsang:2013,Hoang:2020,RasskazovKocsis:2019} explored direct gravitational-wave bremsstrahlung captures from 2-body scattering.  Other potential contributions include the deposition of stellar-mass black holes near a massive black hole by the infall of a globular cluster \citep{ArcaSeddaGualandris:2018}.  We quote a selection of the latest predictions for all of these channels in the executive summary.

Meanwhile, \citet{Bellovary:2016, Bartos:2016, Stone:2016, McKernan:2018} emphasised the important role that an accretion disk in an active galactic nucleus could play in enhancing the rate of binary black hole mergers.   The rich physics of the deep potential well combined with gas interactions can yield new merger channels \citep{Tagawa:2020}.  Compact objects may be captured by the disk or directly form in the disk; in fact, gas may not even be strictly necessary for this, as vector resonant relaxation against the stellar background could naturally lead massive objects to settle into a dis \citep{SzolgyenKocsis:2018}. Migration traps have been suggested as breeding grounds for dynamical interactions that will  enhance the BH+BH merger rate \citep{Bellovary:2016,Yang:2019,Secunda:2019,PengChen:2021}, although there is robust debate in the literature about their efficacy \citep[e.g.,][]{PanYang:2021}. Gas accretion may aid gravitational waves in driving binaries toward merger as well as grow the black hole mass.  Meanwhile, the very high escape velocity in the vicinity of a supermassive black hole ensures that recoil kicks are unlikely to eject merger products, so hierarchical mergers should be more common than in other dynamical environments \citep{Yang:2019,Secunda:2020,Tagawa:2021}.  The rich physics of active galactic nuclei make this channel both promising and complex, with many details such as the feedback from BH-gas interaction and accretion remaining a topic of ongoing work, leading to a broad range of merger rate predictions \citep{Tagawa:2020}.

\subsubsection{Hierarchical 3-body systems}

Another notable dynamical formation channel involves hierarchical triple systems (see \citealt{Naoz:2016} for a review).  In these systems, the transfer of angular momentum back and forth between the inner binary and the wider outer orbit of the third companion can cyclically increase the eccentricity of the inner binary \citep{Lidov:1962,Kozai:1962}.  Once the eccentricity of the inner binary becomes sufficiently high, energy can be dissipated by gravitational waves or tides.  This could drive the inner binary to merger \citep[e.g.,][]{Wen:2003,LiuLai:2018}.   This channel may operate in a number of regimes, from isolated stellar triples \citep{SilsbeeTremaine:2017} to dynamically formed triples in clusters \citep{Antonini:2016} to triples in which the outer companion is a massive black hole in a galactic center \citep{AntoniniPerets:2012,Hoang:2018}.  

Unlike other dynamical formation scenarios, which are most promising for BH-BH mergers, hierarchical triples could be important for NS-NS formation, although predictions are sensitive to assumed NS natal kicks \citep{HamersThompson:2019}.   Extensions to the triple channel include higher multiplicity systems, such as quadruples \citep{Fragione:2020, Hamers:2021,VynatheyaHamers:2021}.  Some authors have associated specific gravitational-wave sources, such as GW190814, with formation in hierarchical triples \citep[e.g.,][]{Lu:2021}.

The triple channel can also operate in conjunction with other channels.  For example, stellar evolution and associated mass loss could force a previously stable stellar triple into instability or drive the inner binary toward merger \citep{PeretsKratter:2012,ShappeeThompson:2013,Stephan:2016}, while chemically homogenous evolution and triple dynamics could produce sequential BH-BH mergers \citep{VignaGomez:2021}.  In fact, it is the combination of triple dynamics with stellar evolution, mass transfer, tides and gravitational waves (as well as the possibility of other dynamical interactions with external perturbers) which makes it particularly challenging to analyse.  The uncertain initial conditions, such as a possible correlation between the mutual inclination, separations and eccentricities of the inner and outer binaries \citep{MoeDiStefano:2017}, query the efficacy of this channel \citep{Toonen:2020} (but see \citealt{Rose:2019}, who find that the observed orbits of massive stars may be consistent with being driven by triple dynamics).  

Field binaries perturbed by flybys from other stars represent a special case on the border between isolated and dynamically formed binaries.  Such flybys can enhance the eccentricity of a wide compact-object binary that evolved as an isolated system, driving it to merge by gravitational-wave emission \citep{Raveh:2022,MichaelyNaoz:2022}.

\subsection{Exotica}

So far, we focussed exclusively on black holes of astrophysical origin, rather than primordial black holes forming from the collapse of early-Universe perturbations.  The latter are potentially very interesting sources of gravitational waves.  \citet{Bird:2016,AliHaimoud:2017,ChenHuang:2018} have argued that a significant fraction of dark matter could be contained in black holes with masses of tens of solar masses, and their mergers could be responsible for some of the gravitational-wave signals observed to date.  However, even allowing for a significant amount of mass in primordial black holes, merger rate predictions are sensitive assumptions about the initial distribution of such sources in binaries or small clusters \citep[e.g.,][]{AliHaimoud:2017, Korol:2019, DeLuca:2020b} as well as possible accretion onto primordial black holes \citep{DeLuca:2020a}.   Meanwhile, some of the scenarios proposed for the formation of BH-BH binaries involve physics beyond the standard model \citep{Sakstein:2020} and cosmological coupling \citep{Croker:2021}.

The discovery of the BH-BH merger GW190521, which left behind a $\sim 150\, M_{\odot}$ remnant \citep{GW190521}, naturally renews interest in the possibility of intermediate mass black hole (IMBH) mergers.  IMBH mergers may be a natural consequence of binary or dynamical evolution \citep[e.g.,][]{AmaroSeoaneSantamaria:2009,McKernan:2012,Belczynski:2014VMS}, and the gravitational waves from mergers of few hundred solar-mass IMBHs could be detected by current instruments \citep{Veitch:2015,Graff:2015,IMBBH:O1}.  Alternatively, stellar-mass compact objects could spiral into IMBHs in clusters.  These intermediate-mass-ratio inspirals are also potentially detectable as gravitational-wave signals \citep{Mandel:2008,Haster:2015IMRI,Haster:2016}.  However, the prevalence of IMBHs remains highly uncertain despite a variety of approaches to their detection \citep[e.g.,][]{Pasham:2014,Paynter:2021}.  The challenges associated with observing IMBHs are discussed by \citet{MillerColbert:2004} and \citet{Greene:2019}.  Given the many orders of magnitude uncertainties, we resist the temptation to review the merger rates of binaries involving IMBHs.  Similarly, we do not specifically discuss hierarchical mergers of stellar-mass black holes or IMBHs as a channel for forming supermassive back holes at high redshifts \citep[e.g.,][]{Volonteri:2010,Kroupa:2020}.  The heaviest merging binaries observed through gravitational waves could represent the tail end of this process of hierarchical massive black-hole formation associated with mergers of ultra-dwarf galaxies \citep{PalmeseConselice:2021}, and many more such mergers may be observable with third-generation gravitational-wave detectors \citep{Gair:2009ET}.

\section{Outlook}\label{sec:outlook}

We reviewed the current estimates of compact-binary merger rates, including both direct observational constraints and theoretical models motivated by observations.  We highlighted some of the key sources of uncertainty in both observations and models.  Of course, these uncertainties can be alternatively  viewed as opportunities.  

The number of observations is growing rapidly.  The number of direct gravitational-wave detections has grown from 0 to 50 between 2015 and 2020.  The range, the number of observed sources, and the measurement accuracy will all increase as the detector sensitivity improves \citep{scenarios}.  Eventually, third-generation gravitational-wave instruments will be able to precisely probe the evolving compact binary merger rate and mass distribution across cosmic time \citep[e.g.,][]{Kalogera:2019,Ng:2020}.  The number and quality of electromagnetic observations are also rapidly growing, again with a very positive outlook, e.g., for kilonovae as new instruments like the Vera C.\ Rubin Observatory come online \citep[e.g.,][]{Chase:2021}.  

These improvements in observations are being matched by a growing sophistication in the models.  On the one hand, increasingly complex models of binary evolution and dynamics are sensitive to a broad range of uncertain assumptions about initial conditions, stellar evolution, mass loss, mass transfer, supernova kicks, and the cosmic star formation history of the Universe.  On the other hand, there is growing evidence that observations of the rates and properties of compact-object mergers, coupled with other observational constraints, will make it possible to resolve these uncertainties and come away with a deep understanding of the physics of compact binary formation \citep[e.g.,][]{MandelOShaughnessy:2010,Stevenson:2015,Barrett:2017FIM,Wong:2021}.

While we focussed on overall rate estimates in this review, specific formation channels are likely to carry distinguishable observational signatures.  We mention two predictions for merging binary black holes by way of example.  Dynamically formed binary black holes should have isotropically oriented spin directions, while BH-BH mergers from isolated binary evolution are expected to have spins preferentially aligned with the orbital angular momentum \citep[e.g.,][]{Farr:2017,Rodriguez:2016spin}.  Is there already evidence for a separate population of dynamically formed systems based on spin-orbit misalignment angles \citep{GWTC2:pop,Roulet:2021,Galaudage:2021}?  Meanwhile, the orbits of most binary black holes are expected to circularise through gravitational-wave emission by the time they approach merger and become detectable with current instruments \citep{Peters:1964}.  But a $\sim 5\%$ fraction of binaries that evolve through binary-single dynamical scatterings may buck the trend, becoming captured at very high eccentricities and small separations through gravitational-wave bremsstrahlung during chaotic interactions \citep{Samsing:2014, Rodriguez:2018}.  Could the tentative evidence for eccentric binary black holes in gravitational-wave data \citep{RomeroShaw:2021} lead to an inferred measurement of the rate of dynamically formed binaries \citep{Zevin:2021}?

Reviewing such a rapidly developing field presents obvious challenges, and we are aware that we have only shown a selection of highlights.  We hope that this is still a useful snapshot of the status of the field against which future rapid progress can be compared.  In fact, we find ourselves in the awkward position of wishing for this review to become obsolete soon after it is written, to be updated as the field moves forward!


\begin{acknowledgements}
\label{sec:acknowledgements}

We are grateful to Iminhaji Ablimit, Manuel Arca Sedda, Chris Belczynski, Edo Berger, Alison Farmer, Saavik Ford, Gabriele Franciolini, Hongwei Ge, Adrian Hamers, Carl-Johan Haster, Ryosuke Hirai, Griffin Hosseinzadeh, Stephen Justham, Bence Kocsis, Pavel Kroupa, Boyuan Liu, Wenbin Lu, Michela Mapelli, Barry McKernan, Selma de Mink, Alex Nitz, Mathieu Renzo, Smadar Naoz, Carl Rodriguez, Navin Sridhar, Thomas Tauris, Ed van den Heuvel, Salvatore Vitale, Tom Wagg, Lev Yungelson, Michael Zevin, and members of Team COMPAS and the WERRD group for discussions and suggestions.  IM carried out parts of this work at the Aspen Center for Physics, which is supported by National Science Foundation grant PHY-1607611, and at MIAPP.  IM acknowledges support from the Australian Research Council Centre of Excellence for Gravitational  Wave  Discovery  (OzGrav), through project number CE17010004.  IM is a recipient of the Australian Research Council Future Fellowship FT190100574.  
\end{acknowledgements}

\section*{Data availability}
The collated data and code to reproduce all figures in this review are publicly available through \citet{ZenodoReview:2021}.

\small
\phantomsection
\addcontentsline{toc}{section}{References}
\bibliography{Mandel}

\end{document}

%% file: table-BNS.tex
\begin{table}
\caption{Summary of local NS-NS coalescence rates, part 1: observations
(\href{https://github.com/FloorBroekgaarden/Rates_of_Compact_Object_Coalescence/blob/main/COC_rates_supplementary_material.pdf}{Supplementary material}
\ and \href{https://zenodo.org/record/5847743}{Zenodo})
}  
\label{table:BNS1}
	\centering
\begin{tabular}{@{}lll@{}}
	\toprule
	\hline \hline
	Source  & Rate & Reference \\ 
	& [Gpc$^{-3}$ yr$^{-1}$] &\\
	\midrule
    %
	\hline \hline
	{\it Observations:} & & \\
	\hline
	Gravitational-wave observations, PDB (ind)  & $44^{+96}_{-34}$ & GWTC-3; \citet{GWTC3:pop}\\
	Gravitational-wave observations, MS & $660^{+1040}_{-530}$ & GWTC-3; \citet{GWTC3:pop}\\
	Gravitational-wave observations, BGP  & $98.0^{+260.0}_{-85.0}$ & GWTC-3; \citet{GWTC3:pop}\\
	\hline
	%
	%
	%
	Short GRBs & $[5, 1800]$ & \citet{Coward:2012}\\
	Short GRBs & $[500, 1500]$ & \citet{Petrillo:2013} \\ 
	Short GRBs & $270^{+1580}_{-180}$ & \citet{Fong:2015} \\	
	 Short GRBs, based on GW170817 & $352^{+810}_{-281}$ & 	\citet{DellaValle:2018}\\
	Short GRBs & $1109^{+1432}_{-657}$ & \citet{Jin:2018}\\
	%
	%
	Short GRBs, based on GW170817 & $ 190^{+440}_{-160}$ &  \citet{Zhang:2018}  \\ 
    Short GRBs, based on GW170817, SWIFT & $160^{+200}_{-100}$ & \citet{Dichiara:2020}\\
    Short GRBs, & $\sim 150$ &  \citet{Zevin:2022sGRB}\\
	\hline
	Kilonovae lower limit & $>8.1$ & \citet{Jin:2016}\\
	Kilonovae, DES, upper limit & $<24000$ & \citet{Doctor:2017}\\
	Kilonovae, PTF, upper limit & $<800$ & 	\citet{Kasliwal:2017}  \\
	Kilonovae, ATLAS, upper limit & $<30000$ & \citet{Smartt:2017}\\
	Kilonovae, DLT40, upper limit & $<99000$ & \citet{Yang:2017}\\
	Kilonovae, ZTF, upper limit & $<900$ & \citet{Andreoni:2021}\\
	\hline
	Galactic pulsar binaries	&$\approx 830^{+2110}_{-680}$ & \citet{OShaughnessyKim:2009}\\
	Galactic pulsar binaries	& $250^{+330}_{-160}$& \citet{Kim:2015}\\
	Galactic pulsar binaries & $ 450^{+290}_{-140}$& \citet{Pol:2020}\\
	Galactic pulsar binaries & $ 370^{+230}_{-100} $& \citet{Grunthal:2021}\\
	%
	%
    %
\end{tabular}
\end{table}

\begin{table}
\ContinuedFloat
  \caption{Summary of local NS-NS coalescence rates, part 2:  model predictions}
  \label{table:BNS2}
	\centering
\begin{tabular}{@{}lll@{}}
\hline 
\hline
	Source  & Rate & Reference \\ 
	& [Gpc$^{-3}$ yr$^{-1}$] &\\      
	\hline \hline
	{\it Models:} & & \\
	\hline
		%
	%
	Isolated binary population synthesis, StarTrack & $[30, 1700]$ &  \citet{OShaughnessy:2009} \\		
	Isolated binary population synthesis, Scenario Machine & $[1050, 3860]$ & 	\citet{LipunovPruzhinskaya:2014}\\	
	Isolated binary population synthesis, Brussels code & $[\leq 1.3, 1800]$ &   \citet{Mennekens:2014}  \\
	Isolated binary population synthesis, StarTrack & $[30, 540]$  & \citet{deMinkBelczynski:2015}\\
	Isolated binary population synthesis, StarTrack & $[52, 162]$&  \citet{Dominik:2014}\\
	Isolated binary population synthesis, BSE & $[240, 1800]$ & \citet{AblimitMaeda:2018} \\
	Isolated binary population synthesis,  StarTrack  & 	$[8, 50]$ 	&\citet{Belczynski:2017}\\
	Isolated binary population synthesis, StarTrack & $[1.5, 631]$ & \citet{Chruslinska:2018}\\		
	Isolated binary population synthesis, MOBSE & $[10, 510]$ &  \citet{GiacobboMapelli:2018}\\	
	Isolated binary population synthesis, StarTrack & $[24, 68]$    & 	\citet{Klencki:2018}\\
	Isolated binary population synthesis, COMBINE & $[2.7, 159]$ &   \citet{Kruckow:2018}\\
	Isolated binary population synthesis, MOBSE  & $[19, 591]$ &  \citet{MapelliGiacobbo:2018} \\
	Isolated binary population synthesis, COMPAS  & $[61.5, 362]$ & \citet{VignaGomez:2018}\\
	Isolated binary population synthesis, MOBSE & $238 $ & \citet{Artale:2019} \\
	Isolated binary population synthesis,  MOBSE &$ [12, 400]$ & \citet{Baibhav:2019} \\	
	Isolated binary population synthesis, SEVN & $70$ &  \citet{Boco:2019}\\	
	Isolated binary population synthesis,  StarTrack  & $[48, 885]$ & \citet{Chruslinska:2019} \\				
	Isolated binary population synthesis, BPASS & $[339, 2178]$ &   \citet{Eldridge:2019} \\	
	Isolated binary population synthesis, COMPAS  & $[20, 245]$ & \citet{Neijssel:2019}\\	
	Isolated binary population synthesis,  StarTrack  & $[49.3, 524]$ 	&\citet{Belczynski:2020}\\
	Isolated binary population synthesis, MOBSE  & $[20, 640]$  & \citet{GiacobboMapelli:2020}\\	
	Isolated binary population synthesis, MOBSE  & $283^{+97}_{-75}$	& \citet{Santoliquido:2020}\\	
	Isolated binary population synthesis, BPASS  &  $[394, 3190]$ &  \citet{Tang:2020}\\
	Isolated binary population synthesis, COSMIC  & $[600, 8900]$ &\citet{Zevin:2020}\\	
	Isolated binary population synthesis, BSE, & $[0.4, 1404]$ & \citet{Chu:2021} \\
	Isolated binary population synthesis, BPASS  &  $[43, 745]$  &	\citet{Ghodla:2021}\\	
	Isolated binary population synthesis, StarTrack  & $[148, 322]$  &	\citet{Olejak:2021} \\ 
	Isolated binary population synthesis, MOBSE & $[4.3, 1036.8]$ &  \citet{Santoliquido:2021}\\		
	Isolated binary population synthesis, BPASS, & $27$ & \citet{Briel:2022}\\
	Isolated binary population synthesis, COMPAS & $[0.32, 330]$ & \citet{Broekgaarden:2021,Broekgaarden:2022} \\	
	Isolated binary population synthesis, StarTrack  & $[116, 155] $ & \citet{Olejak:2022}\\
%
	%
	%
	%
%
	\hline
	Hierarchical triples, SecularMultiple & $[164, 3793]$ &  	 \citet{HamersThompson:2019} \\	
	Hierarchical quadruples, MSE & $[0.8, 30.2]$ &   \citet{VynatheyaHamers:2021}\\
	\hline		
		Globular cluster dynamics  & $30$ &  \citet{Lee:2009} \\
		Globular cluster dynamics  & $[0.32, 3.2]$   & \citet{Bae:2014} \\
		Globular cluster dynamics & $121$& \citet{Samsing:2014}\\
		Globular cluster dynamics, MOCCA & 	$[0.02, 0.5]$ 	&\citet{Belczynski:2017}\\
		Globular cluster dynamics, CMC & $[0.009, \lesssim25.5]$  & \citet{Ye:2019}  \\
		\hline
		Nuclear star cluster dynamics with SMBH &  $[0.004, 1.4]$ & \citet{AntoniniPerets:2012} \\
		Nuclear star cluster dynamics, with SMBH  & $\lesssim 0.02$ & \citet{PetrovichAntonini:2017}  \\
	Nuclear star cluster dynamics  & 	$[0.007, 0.1]$ & \citet{Belczynski:2017}\\
	Nuclear star cluster dynamics, with SMBH & $[0.06, 0.1] $ & \citet{Fragione:2019}\\
	Nuclear star cluster dynamics, with SMBH & $\lesssim 400 $ & \citet{McKernan:2020}\\
	Nuclear star cluster dynamics, with SMBH  & $\gtrsim [0.15, 0.3]$& \citet{Wang:2020}\\
	%
	%
	\hline
	Young star clusters & $[0.03, 0.15]$ & \citet{Ziosi:2014}\\
	Young/Open star clusters, Nbody7 & $[0.01 - 0.1]  $	& \citet{FragioneBanerjee:2020}\\		
	Young star clusters, MOBSE  & $151^{+59}_{-38}  $	& \citet{Santoliquido:2020}\\	
	\bottomrule
\end{tabular}
\end{table}

%% file: table-BHNS.tex
\begin{table*}
\caption{Summary of local NS-BH coalescence rates
(\href{https://github.com/FloorBroekgaarden/Rates_of_Compact_Object_Coalescence/blob/main/COC_rates_supplementary_material.pdf}{Supplementary material}
\ and \href{https://zenodo.org/record/5847743}{Zenodo})
}
\label{table:NSBH}
	\centering
\begin{tabular}{@{}lll@{}}
	\toprule
	Source  & Rate & Reference \\ 		
	& [Gpc$^{-3}$ yr$^{-1}$] &\\
	    %
	\hline \hline
	{\it Observations:} & & \\
	\hline
	Gravitational-wave observations, PDB (ind)  & $73^{+67}_{-37}$ & GWTC-3; \citet{GWTC3:pop}\\
	Gravitational-wave observations, MS & $49^{+91}_{-38}$ & GWTC-3; \citet{GWTC3:pop}\\
	Gravitational-wave observations, BGP  & $32.0^{+62.0}_{-24.0}$ & GWTC-3; \citet{GWTC3:pop}\\
	%
	    %
	%
	%
	\hline 
	Galactic pulsar binaries	& $\lesssim 1800$&  \citet{Pol:2021} \\
	%
	\hline \hline
	{\it Models:} & & \\
	\hline
	Isolated binary population synthesis, StarTrack & $[10, 280]$     & \citet{OShaughnessy:2009} \\
	Isolated binary population synthesis,  Brussels code & $[0.06, 800]$ & \citet{Mennekens:2014} \\	
	Isolated binary population synthesis, StarTrack & $[9, 115]$       &  \citet{deMinkBelczynski:2015}\\
	Isolated binary population synthesis,  StarTrack & $[0.04, 20]$   & \citet{Dominik:2014}\\		
	Isolated binary population synthesis, BSE & $[1, 160]$ & \citet{AblimitMaeda:2018} \\
	Isolated binary population synthesis,  MOBSE & $[5, 780]$ & \citet{GiacobboMapelli:2018}\\
	Isolated binary population synthesis, StarTrack & $[13.3, 26.7]$ &  \citet{Klencki:2018}\\	
	Isolated binary population synthesis,  COMBINE & $[2, 53]$ &  \citet{Kruckow:2018}\\		
	Isolated binary population synthesis, MOBSE  & $[9, 115]$ &  \citet{MapelliGiacobbo:2018} \\		
	Isolated binary population synthesis, MOBSE & $ 78 $ & \citet{Artale:2019} \\
	Isolated binary population synthesis,  MOBSE &$ [4, 37] $ & \citet{Baibhav:2019} \\
	Isolated binary population synthesis, SEVN & $20$&  \citet{Boco:2019}  \\
	Isolated binary population synthesis, StarTrack  & $[5, 230]$ &  \citet{Chruslinska:2019} \\
	Isolated binary population synthesis, BPASS & $[209, 269]$ &   \citet{Eldridge:2019} \\		
	%
	%
	Isolated binary population synthesis, COMPAS  &  $[19, 204]$ &  \citet{Neijssel:2019}\\		
	Isolated binary population synthesis, StarTrack  &  $[0.48, 297]$ & \citet{Belczynski:2020}\\
	Isolated binary population synthesis, MOBSE  & 	$[6, 80]$ 	& \citet{GiacobboMapelli:2020}\\
	Isolated binary population synthesis, MOBSE  & $49^{+48}_{-34}  $	& \citet{Santoliquido:2020}\\	
	Isolated binary population synthesis, BPASS  &  $[58, 6225]$ 	&  \citet{Tang:2020}\\	
	Isolated binary population synthesis, COSMIC  & $[3.7, 1100]$  &\citet{Zevin:2020}\\
	Isolated binary population synthesis, BPASS  &  $[8.7, 498]$  &	\citet{Ghodla:2021}\\
	Isolated binary population synthesis, StarTrack  &  $[4, 16]$  &	\citet{Olejak:2021} \\ 
	Isolated binary population synthesis, POSYDON  &  $[5.7, 77]$ &  \citet{RomanGarza:2021}\\		
	Isolated binary population synthesis,  MOBSE  & $[1.8, 128]$ & \citet{Santoliquido:2021}\\
	Isolated binary population synthesis, BSE  & $[10, 72]$ & \citet{Shao:2021} \\
	Isolated binary population synthesis, BPASS, & $27$ & \citet{Briel:2022}\\
	Isolated binary population synthesis, COMPAS  & $[2.2, 830]$ &	\citet{Broekgaarden:2021,Broekgaarden:2022} \\
	Isolated binary population synthesis, StarTrack  & $[4, 27] $ & \citet{Olejak:2022}\\
%
%
		%
	%
	%
	%
	%
	%
	%
    %
    %
	%
%
	\hline
	Chemically homogeneous evolution, MESA & $[0.02, 0.2]$ & \citet{Marchant:2017}\\
	\hline
	Population III stars & $[0.0002, 0.016]$ & \citet{Belczynski:2017popIII}   \\
	\hline
	Hierarchical triples & $[1.9\times10^{-4}, 22]$ & \citet{FragioneLoeb:2019a,FragioneLoeb:2019b}\\
	Hierarchical triples, SecularMultiple  & $[345, 680]$ &  	 \citet{HamersThompson:2019} \\	
	Hierarchical triples in young star clusters, OKINAMI, & $ [0.04, 0.34]$ & \citet{Trani:2021} \\	
	Hierarchical quadruples, MSE & $[5.7, 57.1]$ &   \citet{VynatheyaHamers:2021}\\
	\hline
	Wide isolated binaries with flybys &$40$ & \citet{MichaelyNaoz:2022}\\
	\hline
	Globular cluster dynamics & $[0.01, 0.12]$ &  \citet{Clausen:2013} \\
	Globular cluster dynamics, ARCHAIN & $\lesssim 0.1$ &  \citet{ArcaSedda:2020} \\
	Globular cluster dynamics, CMC & $[0.009, \lesssim 5.5]$ & \citet{Ye:2019}  \\
	\hline
	Nuclear star cluster dynamics, with SMBH  & $[0.02, 0.4]$ & \citet{PetrovichAntonini:2017} \\
	Nuclear star cluster dynamics, with SMBH & $[0.17, 0.52] $ & \citet{Fragione:2019} \\
	Nuclear star cluster dynamics, with SMBH &  $\gtrsim [2, 5]$ &\citet{Stephan:2019}\\
	Nuclear star cluster dynamics,  & $\lesssim 0.01$ &  \citet{ArcaSedda:2020} \\
	Nuclear star cluster dynamics, with SMBH & $\lesssim 300 $ & \citet{McKernan:2020}\\
	Nuclear star cluster dynamics, with SMBH  & $\gtrsim [0.15, 0.3]$& \citet{Wang:2020}\\
	\hline
	Young star clusters, starlab & $\lesssim 0.1$ &  \citet{Ziosi:2014}\\
	Young/Open star clusters, Nbody7 & $\lesssim 3 \times 10^{-3} $	& \citet{FragioneBanerjee:2020}\\	
	Young star clusters, MOBSE & $\lesssim 28$ &  \citet{Rastello:2020} \\
	Young star clusters, MOBSE  & $41^{+33}_{-23}$ 	& \citet{Santoliquido:2020}\\	
	Young star clusters  & $ [0.04, 36.6]$ &  \citet{ArcaSedda:2021} \\
	%
\bottomrule
\end{tabular}
\end{table*}

%% file: table-BBH.tex
\begin{table}
  \caption{Summary of local BH-BH coalescence rates, part 1: observations and isolated binary evolution model predictions
(\href{https://github.com/FloorBroekgaarden/Rates_of_Compact_Object_Coalescence/blob/main/COC_rates_supplementary_material.pdf}{Supplementary material}
\ and \href{https://zenodo.org/record/5847743}{Zenodo})
  }
  \label{table:BBH1}
	\centering
\begin{tabular}{@{}lll@{}}
\hline 
\hline
	Source  & Rate & Reference \\ 
	& [Gpc$^{-3}$ yr$^{-1}$] &\\
	\hline \hline
	{\it Observations:} & & \\
	\hline
	Gravitational-wave observations, PDB (ind)  & $ 22^{+8.0}_{-6.0} $ & GWTC-3; \citet{GWTC3:pop}\\
	Gravitational-wave observations, MS & $37^{+24}_{-13}$ & GWTC-3; \citet{GWTC3:pop}\\
	Gravitational-wave observations, BGP  & $33.0^{+16}_{-10.0}$ & GWTC-3; \citet{GWTC3:pop}\\
	Gravitational-wave observations, $z$-dependent & $17^{+10}_{-6.7}$ & GWTC-3; \citet{GWTC3:pop}\\
	%
	
	\hline \hline
	{\it Models:} & Rate  &  \\
	\hline
	Isolated binary population synthesis, StarTrack & $[2, 40]$ &  \citet{OShaughnessy:2009} \\ 
	Isolated binary population synthesis, Brussels code & $[\leq 96, 1140]$ &   \citet{Mennekens:2014}\\
	Isolated binary population synthesis, StarTrack & $[14, 2500]$ & \citet{deMinkBelczynski:2015}\\	%
	Isolated binary population synthesis, StarTrack & $[0.5, 221]$& \citet{Dominik:2014}\\	
	Isolated binary population synthesis, BSE & $850$ & \citet{Lamberts:2016}\\
	Isolated binary population synthesis, Scenario Machine & $100$ & \citet{Lipunov:2016}\\
	Isolated binary population synthesis, MOBSE &  $[20, 572]$  &  \citet{Mapelli:2017} \\
	Isolated binary population synthesis, BSE & $[20, 320]$ & \citet{AblimitMaeda:2018} \\
	Isolated binary population synthesis, StarTrack  & $[32, 1072]$ & \citet{Chruslinska:2018}\\	
	Isolated binary population synthesis, MOBSE & $[43, 1500]$ &  \citet{GiacobboMapelli:2018}\\
	Isolated binary population synthesis, StarTrack & $[89, 203]$ &	\citet{Klencki:2018}\\
	Isolated binary population synthesis, COMBINE & $[0.6, 109]$ &  \citet{Kruckow:2018}\\
	Isolated binary population synthesis, MOBSE  & $[146, 240]$ &  \citet{MapelliGiacobbo:2018} \\
	Isolated binary population synthesis, MOBSE & $142$ & \citet{Artale:2019}  \\	
	Isolated binary population synthesis,  MOBSE &$ [30, 60]$ & \citet{Baibhav:2019} \\	
	Isolated binary population synthesis, StarTrack & $[12, 1072]$ &  \citet{Chruslinska:2019} \\	
	Isolated binary population synthesis, BPASS &  $[56, 174]$. &   \citet{Eldridge:2019} \\	
	Isolated binary population synthesis, COMPAS  & $[59, 1157]$ & \citet{Neijssel:2019}\\ 
	Isolated binary population synthesis, SEVN & $90$ & \citet{Spera:2019}\\
	Isolated binary population synthesis, StarTrack & $[1.24, 1368]$ &  \citet{Belczynski:2020}\\	
	Isolated binary population synthesis, MOBSE  & 	$[43, 160]$
	&  \citet{GiacobboMapelli:2020}\\	
	Isolated binary population synthesis, MOBSE  & $[6, 37]$  	& \citet{Mapelli:2020}\\		
	Isolated binary population synthesis, COMPAS & $[51, 87]$  & \citet{Riley:2020}  \\	
	Isolated binary population synthesis, POSYDON  &  $[70, 203]$ &  \citet{RomanGarza:2021}\\
	Isolated binary population synthesis, MOBSE  & $ 50^{+71}_{-37}   $  	& \citet{Santoliquido:2020}\\	
	Isolated binary population synthesis, BPASS  &  $[10, 219]$ & \citet{Tang:2020}\\	
	Isolated binary population synthesis, COSMIC  & $[84, 6900]$ &\citet{Zevin:2020}\\
	Isolated binary population synthesis, POSYDON  &  $[39, 170]$ &  \citet{Bavera:2020}\\	
	Isolated binary population synthesis, BPASS  &  $[31, 873]$  &	\citet{Ghodla:2021}\\	
	Isolated binary population synthesis, StarTrack  & $[18, 89]$  &	\citet{Olejak:2021} \\ 
	Isolated binary population synthesis, MOBSE  & $[10, 105.4]$& \citet{Santoliquido:2021}\\	
	Isolated binary population synthesis, BSE  & $[43, 76]$ & \citet{Shao:2021} \\
	Isolated binary population synthesis, BPASS, & $6.5$ & \citet{Briel:2022}\\
	Isolated binary population synthesis, COMPAS & $[3.8, 810]$ & \citet{Broekgaarden:2021,Broekgaarden:2022} \\	
	Isolated binary population synthesis,  SeBa &  $[19.1, 101.7]$ & \citet{DorozsmaiToonen:2022}\\
	Isolated binary population synthesis, StarTrack  & $[43, 102] $ & \citet{Olejak:2022}\\
	%
	%


	%
	%
	%
	%
%
	%
		\hline
	%
	Chemically homogeneous evolution & $[2, 80]$& \citet{MandelDeMink:2016}\\
	Chemically homogeneous evolution, MESA & $[0.7, 16]$ & \citet{Marchant:2016}   \\
	Chemically homogeneous evolution, MESA  & $[5.8, 7]$ & \citet{duBoisson:2020}  \\
	Chemically homogeneous evolution, COMPAS & $[4, 32]$  & \citet{Riley:2020}  \\
	%
		\hline
	%
	Population III stars & $[12, 25]$ & \citet{Kinugawa:2014} \\
	Population III stars, StarTrack & $[0.016, 1.9]$ & \citet{Belczynski:2017popIII}   \\	
	Population III stars, BSE & $[0.38, 2.9]$ & \citet{Hijikawa:2021} \\
	Population III stars, BSE & $ [3.34,21.2]$ & \citet{Kinugawa:2020} \\
	Population III stars 	in Nuclear Star Clusters & $[0.02, 0.6]$ & 	\citet{LiuBromm:2021}\\
	Population III stars, BSE & $0.1$ & \citet{Tanikawa:2021} \\
	\hline 
\end{tabular}
\end{table}

\begin{table}
\ContinuedFloat
  \caption{Summary of local BH-BH coalescence rates, part 2: dynamical formation models and primordial black hole predictions}
  \label{table:BBH2}
	\centering
\begin{tabular}{@{}lll@{}}
\hline 
\hline
	Source  & Rate & Reference \\ 
	& [Gpc$^{-3}$ yr$^{-1}$] &\\
	\hline \hline
	Hierarchical triples, Rebound package 	& $[0.14, 6.3]$  &	 \citet{SilsbeeTremaine:2017} \\	
	Hierarchical triples, TRES   & $[0.3, 2.5]$  &  	\citet{Antonini:2017} \\	
	Hierarchical triples & $[2, 25]$  & \citet{RodriguezAntonini:2018}\\
	Hierarchical triples in globular clusters, CMC 	& $ [\gtrsim 0.35 , 1]$  &  \citet{Martinez:2020}	\\
	Hierarchical triples in young star clusters & $ [0.2, 1.44]$ & \citet{Trani:2021} \\	 
	Hierarchical quadruples, MSE & $ [8.5, 39.8] $ &   \citet{VynatheyaHamers:2021}\\
	Hierarchical triples, MOBSE and TSE & $[12.5, 95.2]$ & \citet{Stegmann:2022}\\
	%
	%
	%
	%
	\hline
	Wide isolated binaries with flybys, COMPAS & $[1, 20]$ & \citet{Raveh:2022}\\
	\hline
	%
	%
	%
	%
	%
	%
	Globular cluster dynamics  & $[7.25, 29]$  & \citet{Bae:2014} \\
	Globular cluster dynamics, CMC  & $[3.8,  13]$ & \citet{Rodriguez:2015}\\
	Globular cluster dynamics  & $5$ & \citet{AntoniniRasio:2016}  \\
  	Globular cluster dynamics, CMC  & $[2, 20]$ & \citet{Rodriguez:2016big}  \\
	Globular cluster dynamics, MOCCA & $[5.4, 30]$ & \citet{Askar:2016} \\
	Globular cluster dynamics & $[13, 57]$ &  \citet{Fujii:2017}\\
	Globular cluster dynamics, Nbody6  & $[6.5, 26]$ & \citet{Park:2017}  \\
	Globular cluster dynamics   & $[4, 60]$ & \citet{FragioneKocsis:2018} \\ 
	Globular cluster dynamics   & $[0.8, 20]$ & \citet{Hong:2018} \\
	Globular cluster dynamics, CMC  & $[4, 18]$ & \citet{RodriguezLoeb:2018} \\
	Globular cluster dynamics  & $ \approx 6$ & \citet{Choksi:2019} \\	
	Globular cluster dynamics, cBHBd  & 	$[0.2, 50]$ & \citet{AntoniniGieles:2020} \\
	Globular cluster dynamics, CMC & $[9, 30]$ &  \citet{Kremer:2020} \\
	Globular cluster dynamics, MOBSE  & 	$[0.8, 7]$ &  \citet{Mapelli:2020}\\		
	\hline 
	Nuclear star cluster dynamics, without SMBH  & $[1, 10]$ & \citet{MillerLauburg:2008}\\
	Nuclear star cluster dynamics, with SMBH & $[0.002, 0.6]$  & \citet{AntoniniPerets:2012} \\
	Nuclear star cluster dynamics, without SMBH  & 	$1.5$  &  \citet{AntoniniRasio:2016}\\	
	Nuclear star cluster dynamics:, with SMBH & $\sim 1.2$ & \citet{Bartos:2016}\\
	Nuclear star cluster dynamics, with SMBH  & $[0.6, 15]$ & \citet{PetrovichAntonini:2017} \\
	Nuclear star cluster dynamics, with SMBH & $\sim 3$ & \citet{Stone:2016} \\
	Nuclear star cluster dynamics, with SMBH &  $[0.01, 0.4]$  & \citet{Hamers:2018} \\
	Nuclear star cluster dynamics, with SMBH & $[1, 3]$ & \citet{Hoang:2018}  \\
	 Nuclear star cluster dynamics, with SMBH & $[0.17, 0.52] $ & \citet{Fragione:2019}\\
	Nuclear star cluster dynamics, with SMBH & $[0.002, 0.04]$ & \citet{RasskazovKocsis:2019}\\
	Nuclear star cluster dynamics, with SMBH &  $\gtrsim [7, 15]$ &\citet{Stephan:2019}\\
	Nuclear star cluster dynamics with SMBH & $[3.3, 8.6]$ & \citet{ArcaSedda:2020nuc}\\
	Nuclear star cluster dynamics with SMBH & $[0.002, 18]$ & \citet{Grobner:2020}\\
	Nuclear star cluster dynamics, with SMBH & $[0.02, 60]$ & \citet{Tagawa:2020}\\
	Nuclear star cluster dynamics, with SMBH  & $[0.1, 1.6]$. & \citet{Yang:2020} \\
	Nuclear star cluster dynamics, with SMBH  & $[6, 20]$ & \citet{FordMcKernan:2021}\\
	Nuclear star cluster dynamics &  $[0.09, 10]$  &  \citet{Mapelli:2020}\\	
	Nuclear star cluster dynamics, with SMBH &  $\gtrsim [0.3, 5]$ & \citet{Wang:2020}\\
	Nuclear star cluster dynamics, with SMBH & $ [0.083, 14]$ & \citet{Boekholt:2022} \\
	Nuclear star cluster dynamics & $ \lesssim 1 $ & \citet{Codazzo:2022}\\
	\hline 
	Young star clusters, starlab & 	$\lesssim 1.5$  & \citet{Ziosi:2014}  \\
	Young star clusters,  starlab & 	$\gtrsim  1$ &   \citet{Mapelli:2016}  \\
	Open star clusters, NBODY7   & 	$\lesssim 2$ &  \citet{Rastello:2019}\\	
	Young star clusters, NBODY6 & $\lesssim [55, 110] $ &  \citet{DiCarlo:2020} \\
	Open star clusters, NBODY6   & 	$[35, 70]$ & \citet{Kumamoto:2020}\\
	Young star clusters, MOBSE  & $64^{+34}_{-20}  $	& \citet{Santoliquido:2020}\\	
	Young and open star clusters, NBODY7   & 	$[0.5, 37.9]$ & \citet{Banerjee:2021} \\
	Young star clusters, MOBSE  & 	$[0.1, 18]$ &  \citet{Mapelli:2020}\\	
	Young star clusters, NBODY6 MOBSE & $ 88^{+34}_{-26}$ &   \citet{Rastello:2021}\\
	\hline 
	Primordial binaries  &$ [0.02, \lesssim3]$ & \citet{Bird:2016}  \\
	Primordial binaries & $\lesssim [0.2, 10^5]$ & \citet{AliHaimoud:2017}\\
	Primordial binaries &  $\lesssim [6, 10^4]$ & \citet{Raidal:2019}\\
\bottomrule
\end{tabular}
\end{table}